\documentclass[prd,aps,superscriptaddress,floatfix,nofootinbib,eqsecnum,twocolumn]{revtex4-2}

\pdfoutput=1

\usepackage{amsfonts}
\usepackage{amsmath}
\usepackage{amssymb}
\usepackage{bm}
\usepackage{dcolumn}
\usepackage{graphicx}   
\usepackage[latin1]{inputenc}
\usepackage{latexsym}
\usepackage{rotating}   
\usepackage{hyperref}
\usepackage{subfigure}
\usepackage{color}
\usepackage{changes}
\usepackage{verbatim}
\usepackage{comment}
\usepackage{empheq}
\usepackage{physics}
\usepackage{amsmath,latexsym}
\usepackage{float}
\usepackage{soul}
\usepackage{ulem}
\usepackage{todonotes}
 
 \newcommand{\bq}{\begin{equation}}
 \newcommand{\eq}{\end{equation}}
 \newcommand{\bqn}{\begin{eqnarray}}
 \newcommand{\eqn}{\end{eqnarray}}
 \newcommand{\nb}{\nonumber}
 \newcommand{\lb}{\label}

\begin{document}

\title{Quantum geometric formulation of Brans-Dicke theory for Bianchi I spacetime}

\author{Manabendra Sharma}\email{sharma.man@mahidol.ac.th}
\affiliation{Centre for Theoretical Physics and Natural Philosophy, Nakhonsawan Studiorum for Advanced
Studies, Mahidol University, Nakhonsawan 60130, Thailand}
\affiliation{Inter University Centre for Astronomy and Astrophysics, Post Bag 4, Pune 411007, India}

\author{Gustavo S. Vicente}\email{gustavo@fat.uerj.br}
\affiliation{Faculdade de Tecnologia, Universidade do Estado do Rio de Janeiro,
27537-000 Resende, RJ, Brazil}

\author{Leila L. Graef}\email{leilagraef@id.uff.br}
\affiliation{Instituto de Fi\'sica, Universidade Federal Fluminense, 24210-346 Niteroi, RJ, Brazil}
\author{Rudnei O. Ramos}\email{rudnei@uerj.br}
\affiliation{Departamento de Fi\'sica Te\'orica, Universidade do Estado do Rio de Janeiro,
20550-013 Rio de Janeiro, RJ, Brazil}

\author{Anzhong Wang}\email{Anzhong$_$Wang@baylor.edu}
\affiliation{GCAP-CASPER, Department of Physics and Astronomy, Baylor University,
Waco, Texas 76798-7316, USA}

\date{\today}
\begin{abstract}

This paper investigates Bianchi I spacetimes within the Jordan frame of Brans-Dicke theory, incorporating the framework of effective loop quantum gravity. After developing general formulas, we analyze the robustness of classical singularity resolution due to quantum geometric effects using two common quantization schemes. We then compare the resulting physical properties. We find that both schemes replace classical singularities with regular quantum bounces. Notably, in contrast to similar studies based on general relativity, we find that all three directional scale factors of the Bianchi I spacetimes increase and  {after the quantum bounce they reach values similar to their initial values}, leading to a merging with classical spacetimes in both schemes.

\end{abstract}

\maketitle 

\section{Introduction}
\label{intro}

Loop quantum gravity (LQG) is a candidate of quantum gravity theory,
which takes the premise of gravity as a manifestation of geometry of
spacetime and  systematically constructs a  theory of quantum
Riemanian geometry with rigor (see, e.g., Ref.~\cite{Ashtekar:2021kfp}
for a recent review).  LQG stands out as a nonperturbative background
independent approach to quantize gravity~\cite{Thiemann:2007pyv,Rovelli:2008zza,Bojowald:2010qpa,Gambini:2011zz,Rovelli:2014ssa}.  At
its depth, this theory brings out a fundamental discreteness at Planck
scale wherein the underlying geometric observables, such as areas of
physical surfaces and volumes of physical regions, are discrete in
nature~\cite{Thiemann:2002nj,Ashtekar:2004eh,Rovelli:2004tv,Ashtekar:1994wa}.
At present, the studies in cosmology and black holes provide ones of
the major avenues for applying and testing the ideas of  the theory.  
The absence of the possibility to design a table top
experiment, other than thought experiment, can be easily understood
from the energy scale involved in LQG, which is 
 by far  beyond the reach of
present  technology.  However, it is feasible to apply LQG against
observational physics in the context of quantum  cosmology.
In fact, the early Universe  provides  a natural  laboratory
for this purpose, and it is very interesting  and important to investigate the 
Planckian physics of  the very early  Universe  in
the framework of LQG.

Loop quantum cosmology (LQC) is an application of LQG techniques to
the symmetry reduced spacetimes ~\cite{Ashtekar:1997fb,Ashtekar:2011ni,Li:2021mop,Barca:2021qdn,Li:2023dwy,Agullo:2023rqq}.
In LQC the big bang singularity
is resolved in the sense that it is replaced by a  quantum bounce,
across which physical observables, such as energy density
and curvatures, which diverge at the big bang singularity in classical
general relativity (GR), now all remain finite.  This nondivergent
behavior owes to the fact that 
 in LQG the area operator
has a discrete spectrum of eigenvalues with a nonzero area gap  
\cite{Thiemann:2002nj,Ashtekar:2004eh,Rovelli:2004tv,Ashtekar:1994wa}. 
It is this  nonzero gap that prevents the formation of spacetime singularities
\cite{Bianchi:2008es,Ma:2010fy,Bojowald:2005epg}. This is quite similar to 
the simple harmonic oscillator, whose ground state  energy is not
zero quantum mechanically, although classically it does.
Remarkably, LQC
produces a contracting Friedmann-Lema\^itre-Robertson-Walker (FLRW)
Universe that bounces back to an expanding one, thus, avoiding the
occurrence of a  singularity.  This is achieved without adding any
nontrivial matter component, unlike in the case of classical or matter
bounces (see, e.g., Refs.~\cite{Brandenberger:2012zb,Lehners:2008vx,Cai:2013vm}).
This quantum bounce occurs purely due to quantum geometric effects,  a
novel repulsive effective force that is manifest in the quantum
corrected {}Friedmann and Raychaudhuri equations.  
The massless scalar field that acts as a clock in this setting, is identified with the inflaton that drives a nearly exponential expansion of the Universe given by the effective dynamics. The preinflationary background dynamics have been extensively studied in detail for various forms of potentials in the Refs.~\cite{Sharma:2018vnv,Jin:2018wdx,Li:2018fco,Levy:2024naz,Shahalam:2017wba}. 
In addition, 
in all  spacetimes studied so far, which permit
different symmetries, including the
Bianchi  and Gowdy models~\cite{Chiou:2006qq,Chiou:2007sp,Chiou:2007mg,Martin-Benito:2008dfr,Ashtekar:2009vc,Martin-Benito:2009xaf,Ashtekar:2009um,Corichi:2009pp,Garay:2010sk,Wilson-Ewing:2010lkm,Gupt:2011jh,Singh:2013ava,Tarrio:2013ija,Linsefors:2014tna,Corichi:2015ala,Wilson-Ewing:2017vju,Agullo:2020uii,Agullo:2020iqv,McNamara:2022dmf,Motaharfar:2023hil}, 
the singularity is resolved~\cite{Ashtekar:1997fb,Ashtekar:2011ni,Li:2021mop,Barca:2021qdn,Li:2023dwy,Agullo:2023rqq}.

An important issue that appears in the bouncing scenario is the
possible instability in the growth of anisotropic density during the
contracting phase.  As a matter of fact, during this contraction phase, the contribution from
anisotropic stresses to the Friedmann equation grows much faster
than the energy density of the usual fields, like radiation, baryonic
matter and cold dark matter. Once the ratio of the anisotropic
stresses  to the total energy density becomes comparable, it is
not possible any longer to assume that the background spacetime is
approximately isotropic, and this ratio can even become larger than
unity, eventually 
leading to an anisotropic collapse.  When considering
inhomogeneities, the situation is even more sensitive. In particular, when  the ratio
of the anisotropic stresses to the total energy density  is
larger than unity,  it indicates
the onset of the
conjectured Belinski-Khalatnikov-Lifshitz chaotic
instability~\cite{Belinsky:1970ew}.  As is well-known, due to the
high sensitivity of the dynamics to the initial conditions
(neighboring points following very different dynamics), this scenario
can lead to the loss of predictivity.  In this case, the large
anisotropies impact  the density  fluctuations in a highly
inhomogeneous way, spoiling the prediction of a nearly isotropic
cosmic microwave background~\cite{Aluri:2022hzs,Yeung:2022smn}.
This can be avoided, for example, in some situations when the nearly
scale-invariant perturbations are generated after the
bounce~\cite{Agullo:2022klq}.  Furthermore, in the special case of
the Bianchi type I model~\cite{Amirhashchi:2017mur}, it was shown that
isotropization is an attractor, which makes this scenario a natural
choice to investigate viable candidates. This is especially promising
in the framework of LQC, where we can analyze the Bianchi I background in a
bouncing scenario driven by modifications to the Einstein equations in
the high energy regime.  Great efforts have been devoted towards
understanding the quantum corrected dynamics of anisotropic spacetime
starting with the Einstein-Hilbert action 
(see, for example,
Refs.~\cite{Chiou:2006qq,Chiou:2007sp,Chiou:2007mg,Martin-Benito:2008dfr,Ashtekar:2009vc,Martin-Benito:2009xaf,Ashtekar:2009um,Corichi:2009pp,Garay:2010sk,Wilson-Ewing:2010lkm,Gupt:2011jh,Singh:2013ava,Tarrio:2013ija,Linsefors:2014tna,Corichi:2015ala,Wilson-Ewing:2017vju,Agullo:2020uii,Agullo:2020iqv,McNamara:2022dmf,Motaharfar:2023hil} and references therein).  

A natural question to ask is what will happen when one goes beyond
the Einstein-Hilbert action. One of the salient features of GR
is that all the geometrical quantities, such as distance, area and
volume can be constructed from the knowledge of the metric. The metric
and, hence, the geometry of the spacetime, being a dynamical quantity that is
determined by the distribution of matter and energy. These fundamental
properties of the theory remain intact even beyond the minimally coupled
Einstein-Hilbert action.  This motivates us to consider an action with
a nonminimal coupling term and to find the corresponding quantum
corrected dynamics. As an example toward this goal, we consider the
Brans-Dicke theory (BDT).  In particular, the  techniques of LQG are
not  limited to a particular action.  Based on the loop quantization,
some aspects of its implementation in the context of the BDT has been
explored for the FLRW
spacetime~\cite{Zhang:2011vg,Artymowski:2013qua,Han:2015jsa,Han:2019mvj,Song:2020pqm}
and the presence of a quantum  bounce is shown in both the Jordan and Einstein
frames.

One of the prime focus of the present
paper is to formulate the quantum corrected effective dynamics of BDT
in homogeneous and anisotropic Bianchi I spacetimes in the Jordan
frame, whose study is still missing in the literature, as far as we
are aware of.  This is a crucial step towards establishing and
checking the robustness of the major results of loop quantization in
spacetimes  with different symmetries and different theories of
gravity.  In particular, we aim to ultimately check the robustness of
the singularity resolution and the consistency of the effective
dynamics.

This paper is organized as follows.  In Sec.~\ref{HF}, we 
 briefly review the
Hamiltonian formulation of BDT and provide the necessary formulas for the studies of
Bianchi I spacetimes in Jordan's frame of the theory. 
In Sec.~\ref{SR}, we present the symmetry
reduction of the classical Hamiltonian constraint for Bianchi I spacetimes
and write down the corresponding classical dynamical equations.
In Sec.~\ref{QCD}, we study  the quantum corrected dynamics using two
alternative and complementary quantization schemes, 
which have been considered in the literature, and write down the dynamical equations for each of the schemes.
In order to study quantum effects of matter fields, in the effective Hamiltonian, we also include a massless scalar field $\Phi$,
in addition to the Brans-Dicke (BD) scalar field $\phi$. In
Sec.~\ref{dynamics}, we study numerically the dynamics of the BDT
for Bianchi LQC models in both  quantization schemes  and show explicitly that the classical spacetime singularity is indeed
resolved in both schemes and replaced by a quantum bounce.   Our  main conclusions are presented
in Sec.~\ref{DR}. 
Throughout this work, we will adopt the convention
$M_{\rm Pl}=1$ (unless in the cases where possible confusions can raise), where $M_{\rm Pl}\equiv 1/\sqrt{8 \pi G}$ is the reduced Planck
mass. 

\section{Hamiltonian formulation of Brans-Dicke theory}
\label{HF}

To facilitate our investigations to be carried out in the next sections, 
in this section we shall provide a very brief review of BDT and for more details, we refer the readers to
\cite{Zhang:2011vg,Artymowski:2013qua,Han:2015jsa,Han:2019mvj,Song:2020pqm}. The BDT is described by the action
\begin{eqnarray}
S_{\rm BDT}= \frac{1}{2} \int d^4x \sqrt{-g} \left( \phi R -
\frac{\omega}{\phi} g^{\mu \nu}\partial_{\mu} \phi \partial_{\nu} \phi
\right), \label{BDTAction} 
\end{eqnarray}
where $R$ is the Ricci scalar, $g_{\mu \nu}$  the metric and $g$ 
its determinant. The BD scalar field is denoted by $\phi$ and $\omega$ is the coupling constant, which is highly constrained observationally \cite{Zhang:2017sym}. 
%
%
  The present format of covariant formulations of the theory is not suitable
for loop quantization.  In the Lagrangian formulation, space and time are treated
on equal footing, but the construction of the canonical formula
demands splitting of spacetime into space and time.  It is  the
Arnowitt-Deser-Misner 
formalism~\cite{Kiefer:2004xyv,Bojowald:2010qpa,DeWitt:2007mi} with
the $3+1$ decomposition that comes to rescue  the situation by providing a description in terms
of the first-order Hamilton's equations of the phase space variables.
Formally, this can be represented by decomposing the spacetime manifold, $\mathcal{M}$
endowed with a metric, designated by the pair $(\mathcal{M}, g_{\mu
  \nu})$, into $(\Sigma$$,\mathbb{R})$, where $\Sigma$ represents the
spatial three-dimensional hypersurface and $\mathbb{R}$ stands for the
real valued time.  
 
The splitting of the metric into its
intrinsic spatial metric $q_{ab}\; (a, b = 1, 2, 3)$ and the normal vector $n_{\mu}\; (n_{\nu}n^{\nu}=-1)$
to the hypersurface 
%
%
{ naturally introduces the notion of extrinsic curvature, $K_{ab}$, defined in terms of covariant derivative of the normal vector to the hypersurface as projected to the hypersurface \cite{Baez:1995sj}. Let us denote q as }
the determinat of $q_{ab}$ and $N$ denotes the lapse function. 
Then, the  corresponding conjugate momenta  for the BDT action given by
Eq.~(\ref{BDTAction}) are
\begin{eqnarray}
&&p^{ab} \equiv \frac{\delta \mathcal{L}}{\delta \dot{q}_{ab}}=
\frac{\sqrt{q}}{2} \left[ \phi(K^{ab}-Kq^{ab})-\frac{q^{ab}}{N}
  (\dot{\phi}-N^c\partial_c\phi) \right],\label{CP1}\nonumber\\\\  
&&  p_{\phi}
\equiv \frac{\delta \mathcal{L}}{\delta \dot{\phi}}= -\sqrt{q}
\left[ K-\frac{w}{N \phi}(\dot{\phi}-N^c\partial_c
  \phi)\right], \label{CPi} \\ 
&&  p_N(x) \equiv \frac{\delta
  \mathcal{L}}{\delta \dot{N}(x)} =0, \label{CN}\\
&& p_a(x) \equiv
\frac{\delta \mathcal{L}}{\delta \dot{N}^a(x)}=0,\label{CNa}
\end{eqnarray}
{}{ where the Lagrangian density $\mathcal{L}$ is given by \cite{Olmo:2011fh}}
\begin{eqnarray}
    \mathcal{L} &=& \frac{\sqrt{q}}{2\kappa^2} \Bigg[N\phi\left( \mathcal{R}+(K^{ab}K_{ab}-K^2) \right) +2 q^{ab}D_aND_b{\phi} \nonumber\\
    &-& \frac{\omega}{N\phi} \left(N^2 q^{ab}D_{a}{\phi}D_{b}\phi  -(\dot{\phi}-N^aD_a{\phi})^2 \right)\nonumber\\
    &-&2K(\dot{\phi}-N^aD_a\phi)\Bigg].
\label{Lagrangian}
\end{eqnarray}
{Here the trace of the extrinsic curvature tensor is $K\equiv K^{a}_a=q^{ab}K_{ab}$ and the spatial covariant derivative is defined as $D_a\equiv q^{b}_a \nabla_b$.} 
{}For a given Lagrangian, the Hamiltonian can be obtained by the
Legendre transformation,
\begin{eqnarray} 
    H_{\rm total}&=& \int_{\Sigma} d^3x \sqrt{q} (\dot{q}_{ab}p^{ab}+
    \dot{\phi} p_{\phi}- \mathcal{L}(q_{ab}, \dot{q}_{ab})),\;\;\;
\end{eqnarray}
which can be cast in the form
\begin{eqnarray} 
    H_{\rm total}&=& 
\int_{\Sigma} d^3x\, N^a \left[ -2 D^bp_{ab}+p_{\phi} \partial_a
  \phi\right]
\nonumber \\
&+& \int_{\Sigma} d^3x\, N \left\{ \frac{2}{\sqrt{q}\phi}
\left[ p_{ab}p^{ab}-\frac{1}{2}p^2 + \frac{(p-p_{\phi}
    \phi)^2}{2(3+2\omega)}\right] \right.
\nonumber\\ &+&\left.\frac{\sqrt{q}}{2} \left[-\phi R +
  \frac{w}{\phi}(D_a \phi)D^a\phi + 2D_aD^a \phi
  \right]\right\},\nonumber\\
\end{eqnarray}
In terms of the Asthekar variables, with the densetized
triads $E^a_i\equiv \sqrt{e} e^a_i $, curvature tensor $F^i_{ab} \equiv
\partial_{[a}A^i_{b]} + \epsilon^i_{kl}A^k_lA^k_b $, and the Yang-Mills connection $A^i_a \equiv
\Gamma^i_a + \gamma K^i_a$, the constraints of the BDT  in the Yang-Mills phase space read
\begin{eqnarray}
\mathcal{G}_i &=& D_a E^a_i \approx 0,
\label{GaussConstraint} \\
\mathcal{V}_a &=&\frac{1}{ \gamma} F^i_{ab} E^b_i + p_{\phi}\partial_a \phi\approx 0,
\\ \mathcal{H}&=& \frac{\phi}{2} \left[ F^j_{ab} -\left(
  \gamma^2\ +\frac{1}{\phi^2}\right)\epsilon_{jmn}\tilde{K}^m_a
  \tilde{K}^n_b\right]\frac{\epsilon_{jkl}E^a_kE^b_l}{\sqrt{q}}
\nonumber\\  &+&\frac{1}{3+2\omega}\left[ \frac{(\tilde{K}^i_a
    E^a_i)^2}{ \phi \sqrt{q}} + 2\frac{\tilde{K}^i_aE^a_i}{
    \sqrt{q}}p_{\phi} + \frac{p_{\phi}^2 \phi}{\sqrt{q}} \right] \nonumber\\
    &+& \frac{\omega}{2\phi}\sqrt{q}(D_a \phi)D^a\phi + \sqrt{q}D_aD^a\phi\approx 0,
\nonumber \\
\label{Scalar Constraint}
\end{eqnarray}
which are  often referred to as the  Gauss, diffeomorphism and scalar constraints,
respectively. In the above expressions, $\gamma$ is the Barbero-Immirzi parameter and
$\tilde{K}^i_a= e^b_i \tilde{K} K_{ab}$  with
\begin{eqnarray}
    \tilde{K}^{ab}= \phi K^{ab} + \frac{q^{ab}}{2N}(\dot{\phi}-
    N^c\partial_c \phi).
\end{eqnarray}
In Eqs.~(\ref{GaussConstraint})-(\ref{Scalar Constraint}), the symbol
``$\approx 0$" indicates that the constraints hold only on the shell
(hypersurface).  Thus, the total Hamiltonian of  BDT can be expressed as
\begin{eqnarray}
C_{\rm total} = \int d^3x  ( N^a \mathcal{V}_a + N
\mathcal{H}+\mathcal{G}_i N^i).
\end{eqnarray}

\section{Symmetry Reduction of Classical BDT for Bianchi I Spacetimes}\label{SR}

The four-dimensional Bianchi I spacetimes described by the metric
\bqn
 \label{eq3.1}
 ds^2 &=& -N^2dt^2 + q_{ab}dx^adx^b \nb\\
 &=&   -N^2dt^2 + \sum_{I = 1}^{3}{a_I^{2} \left(dx_I\right)^2}, 
\eqn
have a topology of  $\Sigma
\times \mathbb{R}$, where $N$ and $a_I$ are functions of $t$ only,  $x^{\mu} \equiv (t, x^a)$ and $x^a \equiv (x, y, z)$. Thus, the   intrinsic metric $q_{ab} = 
\text{diag}(a_1^{2}, a_2^{2}, a_3^{2})$ defined on
the spatial manifold $\Sigma$ is homogeneous with three-dimensional
symmetry group that is simply transitive.  The symmetry of the spatial
hypersurface is determined by the Killing vectors.  Essentially, the
Killing vector fields give the direction along which  the
metric does not change under the Lie dragging~\cite{Wilson-Ewing:2017vju}. 

\subsection{Classical Hamiltonian} 

Considering a fiducial cell with lengths $L_1,\; L_2$ and $L_3$ along the three spatial directions 
\cite{Ashtekar:2009vc},
\bqn
 \label{eq3.2}
 ds_o^2 \equiv  {}^{o}q_{ab}dx^adx^b = dx_1^2 + dx_2^2 + dx_3^2, 
\eqn
its triads and cotriads, denoted, respectively,  by
$\mathring{e}^a_i$ and $\mathring{e}^i_a$, satisfy the relations
$\mathring{e}^a_i \mathring{e}^i_b = \delta^a_b$ and $\mathring{e}^a_i
\mathring{e}^j_a = \delta^j_i$. 
{Without loss of the generality, one often sets $L_1 = L_2 = L_3 \equiv l_0$, which will be adopted in this paper.}
The geometry of the Bianchi I spacetimes is one of the simplest types of
Bianchi Class A models \cite{Jantzen:2001me}, such that it admits the Universe to have
different scale factors in each principal direction, but with zero
spatial curvature.  

 {The introduction of fiducial triads trivializes
the gauge freedom corresponding to the Gauss and the spatial diffeomorphisms
constraints}  and they commute with the
Killing vector fields denoted by $\mathring{\xi}^a_i$, satisfying the relations
$[\mathring{e}_i, \mathring{\xi}_j]^a= 0$.
The underlying structure of the spatial hypersurface is dictated by
the commutator relations, given by $[\mathring{e}_i, \mathring{e}_j]^a= C^k_{ij} \mathring{e}^a_k$.
The values of the structure constants $C^k_{ij}$ completely define all
the symmetry of the spatial sector.  The Bianchi I model considered in
this work is specified by  $C^i_{jk}=0$.  To be specific, the Bianchi
I  hypersurface $\Sigma$ is flat. 
Also, the homogeneity implies the contribution to the Hamiltonian from
the spatial derivative is identically zero.  
The diagonal
Bianchi I Universe in terms of the Asthekar variables and densetized
triads are parametrized by
\begin{eqnarray}\label{Para} 
A^i_a = \tilde{c_i} V_o^{-1/3}  \mathring{e}^a_i, \quad
E^b_j = p_j
V_o^{-2/3} \sqrt{q} \mathring{e}^b_j,
\end{eqnarray}
which satisfy the  canonical relations
\begin{eqnarray}
\{A^i_a(x), E^b_j(y)\} = \gamma \delta^i_j \delta^b_a
\delta(x-y),
\end{eqnarray}
where there is no summation over indexes $i$ or $j$,  and the volume of the fiducial cell $V_o$ is given by
$V_o =   l_0^3$.  
Since Bianchi I spacetimes are  homogeneous but anisotropic, 
$\tilde{c_i}$ and $p_j$ depend only on time. 
Also, since there is no
spatial curvature, we have $\Gamma^i_a= 0$. As a result, 
{the only constraint that survives after the adoption of the fiducial triad }
is the Hamiltonian constraint, 
%
%
%
%
which is given by
\begin{eqnarray}
\mathcal{H}  &=& \frac{\phi}{2} \left[ F^j_{ab}-\left(\gamma^2  +
  \frac{1}{\phi^2}\right)\epsilon_{jmn} \tilde{K}^m_a \tilde{K}^n_b
  \right] \frac{\epsilon_{jkl} E^a_k E^b_l}{\sqrt{q}} \nonumber\\ &+&
\frac{1}{3+2\omega} \left[\frac{(\tilde{K}^i_a E^a_i)^2}{ \phi
    \sqrt{q}} + 2\frac{(\tilde{K}^i_a E^a_i) p_{\phi}}{ \sqrt{q}} +
  \frac{p^2_{\phi}\phi}{ \sqrt{q}}\right] \nonumber \\ &+&
\frac{\omega}{2 \phi}\sqrt{q}(D_a \phi)D^a\phi + \sqrt{q} D_a D^a
\phi.
\end{eqnarray}
Since the  Bianchi I spacetimes are spatially flat, $\Gamma^k_a=0$, as pointed out,  we find that 
%
\begin{eqnarray}
    F^j_{ab}= \gamma^2 \epsilon^j_{kl} \tilde{K}^k_a \tilde{K}^l_b.
\end{eqnarray}
We  also notice that for the 
parametrization ~(\ref{Para}), we have
\begin{eqnarray}
&& \epsilon_{jkl} \epsilon_{jmn} E^a_k E^b_l \tilde{K}^m_a \tilde{K}^n_b
=  2 E^a_m E^b_n \tilde{K}^m_{[a} \tilde{K}^n_{b]},\\
\label{1stTerm}
&& E^a_m E^b_n \tilde{K}^m_{[a} \tilde{K}^n_{b]} =
\frac{\mathring{q}}{\gamma^2 l_0^6}
\left(\tilde{c}_1\tilde{c}_2p_1p_2+\tilde{c}_2p_2\tilde{c}_3p_3\right.
\nonumber\\ 
& &~~~~~~~~~~~~~~~~ +\left.\tilde{c}_1p_1\tilde{c}_3p_3\right),\\
&& \tilde{K}^i_a E^a_i =  \frac{\sqrt{\mathring{q}}}{\gamma
  l_0^3}(\tilde{c}_1p_1+\tilde{c}_2p_2+\tilde{c}_3p_3), \quad
\end{eqnarray}
where  $[a,b]\equiv (ab-ba)/2$ denotes the antisymmetrization. 
In the limiting case of $\omega \to \infty $ and $\phi$ being a
constant, the scalar constraint of the BDT gets reduced to the scalar
constraint of Einstein's GR in the vacuum. 

Now, combining all the terms and using that $\sqrt{q}=
\sqrt{\mathring{q}}\sqrt{p_1p_2p_3}/l_0^3$, we finally obtain  
\begin{eqnarray}\label{SymRedHamiltonian}
\mathcal{H}&=&  - \frac{\sqrt{\mathring{q}}}{ l_0^3}
\frac{1}{\sqrt{p_1p_2p_3}} \frac{1}{\gamma^2\phi } \Bigl[
  \tilde{c}_1\tilde{c}_2p_1p_2+\tilde{c}_2p_2\tilde{c}_3p_3+\tilde{c}_1p_1\tilde{c}_3p_3
  \nonumber \\ &-& \frac{1}{\beta} \left(
  \tilde{c}_1p_1+\tilde{c}_2p_2+\tilde{c}_3p_3+ \gamma \pi_\phi \phi
  \right)^2 \Bigr],
\end{eqnarray}
%
{where, in the move to homogeneous variables, we have redefined the conjugate momentum} corresponding to the
scalar field as $p_{\phi} \equiv \pi_{\phi}\sqrt{\mathring{q}}/l_0^3$
and $\beta=3+2\omega$.  The smeared Hamiltonian is obtained by
integrating the symmetry reduced Hamiltonian,
Eq.~(\ref{SymRedHamiltonian}), over the fiducial volume $V_0$, leading
to
\begin{eqnarray}
\!\!\!\!\!  C_\mathcal{H} &=& \int d^3x N \mathcal{H} \nonumber\\ &=&
- \frac{N}{\sqrt{p_1p_2p_3}} \frac{1}{\gamma^2\phi } \Bigl[
  \tilde{c}_1\tilde{c}_2p_1p_2+\tilde{c}_2p_2\tilde{c}_3p_3+\tilde{c}_1p_1\tilde{c}_3p_3
  \nonumber \\ &-& \frac{1}{\beta}  \left(
  \tilde{c}_1p_1+\tilde{c}_2p_2+\tilde{c}_3p_3+ \gamma \pi_\phi \phi
  \right)^2 \Bigr],
\label{FinalHamiltonian}
\end{eqnarray}
where we have used  $\int d^3x\sqrt{\mathring{q}}=l_0^3$.
 
\subsection{Classical dynamics} 

The symplectic structure of the  complete phase space in terms of the
new variables reads
\begin{eqnarray}
\{ \tilde{c}_I, p_J \} &=&  \gamma \delta_{IJ}, \\ \{ \phi, \pi_{\phi}
\}   &=& 1.
\end{eqnarray}
The dynamics of a system is given by the Poisson equation,  $\dot{\mathcal{O}}=
\{\mathcal{O}, C_{H} \}$ 
{for the phase space function} $\mathcal{O}$. 
 Then,  the equations of motion for $c_I$ and its
conjugate variables $p_I$ are
\begin{eqnarray}
\label{cI_EOM}
{\tilde{c}}'_I  &=&  \gamma\,  \delta_{IJ}  \frac{\delta
  C_\mathcal{H}}{\delta p_J}, \\ 
  {p_I}'
\label{pI_EOM}
&=&  - \gamma\, \delta_{IJ} \frac{\delta C_\mathcal{H}}{\delta
  \tilde{c}_J},
\end{eqnarray}
whereas the equations of motion for $\phi$ and its conjugate variable
$\pi_\phi$ are
\begin{eqnarray}\label{phi_EOM}
{\phi}'  &=&  \frac{\delta C_\mathcal{H}}{\delta \pi_\phi}, \\     
\label{pphi_EOM}
{\pi}_\phi'  &=&   - \frac{\delta C_\mathcal{H}}{\delta \phi},
\end{eqnarray}
where a dot denotes the derivative with respect to cosmic time $t$ and a prime denotes derivative with respect to $\tau$ defined as %
\begin{eqnarray}\label{gauge}
d\tau  \equiv N dt = \sqrt{p_1p_2p_3}dt.    
\end{eqnarray}
In this work, we choose $N=\sqrt{p_1p_2p_3}$.

Using the Hamiltonian constraint, Eq.~\eqref{FinalHamiltonian}, we can
now derive the equations of motion. 
In particular, from Eq.~\eqref{cI_EOM}, we obtain the following equation of motion for
the $c_I$ variables,
\begin{eqnarray}
\label{cClassical}
\tilde{c}_I' &=& -\frac{\tilde{c}_I}{\gamma\phi} \Bigl[
  \tilde{c}_Jp_J+\tilde{c}_Kp_K
\nonumber \\
&-& \frac{2}{\beta} \left(
  \tilde{c}_Ip_I+\tilde{c}_Jp_J+\tilde{c}_Kp_K +  \gamma \pi_{\phi}
  \phi  \right) \Bigr], \;\;\; 
\end{eqnarray}
where $J, \; K \not= I$. 

Similarly,
from Eq.~\eqref{pI_EOM}, we obtain the equation of motion for the
conjugate variables $p_I$,
\begin{eqnarray}
\label{pClassical}
{p}_I' &=&   \frac{p_I}{\gamma\phi}  \Bigl[
  \tilde{c_J}p_J+\tilde{c}_Kp_K 
\nonumber \\
&-& \frac{2}{\beta}
  \left(\tilde{c}_Ip_I+\tilde{c}_J p_J +\tilde{c}_Kp_K +\gamma
  \pi_{\phi}\phi\right)\Bigr]. 
\end{eqnarray}
{}Finally, from Eqs.~\eqref{phi_EOM} and~\eqref{pphi_EOM},
respectively, the equations of motion for the BD scalar field $\phi$
and its conjugate momentum $\pi_{\phi}$ are given by
\begin{eqnarray}
\phi' &=&  \frac{2}{\beta\gamma} \left( \tilde{c}_1p_1+
\tilde{c}_2p_2+ \tilde{c}_3p_3 + \gamma\pi_{\phi}\phi \right), 
\label{phidot} \\
\pi_{\phi}'&=&  - \frac{1}{\gamma^2\phi^2} \Bigl\{ \left(
\tilde{c}_1p_1\tilde{c}_2p_2 +  \tilde{c}_2p_2\tilde{c}_3p_3 +
\tilde{c}_1p_1\tilde{c}_3 p_3 \right)   \nonumber\\ &-&
\frac{1}{\beta}\left[ (\tilde{c}_1p_1+ \tilde{c}_2p_2
  +\tilde{c}_3p_3)^2 -(\phi\gamma\pi_{\phi})^2\right] \Bigr\}.
\label{pphidot}
\end{eqnarray}
The set of Eqs.~(\ref{cClassical})-(\ref{pphidot}) completely
specifies the classical dynamics of   the Bianchi-I spacetimes  for BDT.  Note that even at the level of equations of motion, it
is straightforward to see that in the limit $\omega \rightarrow
\infty$ and constant $\phi$, the equations of motion for the
Einstein-Hilbert action are recovered~\cite{Wilson-Ewing:2017vju}.

\section{Loop Quantum Effective Dynamics of Bianchi I Spacetimes in BDT}
\label{QCD}

The loop effective quantum dynamics for homogeneous and isotropic Universe in LQC \cite{Taveras:2008ke} have been studied extensively, 
and now it is well established \cite{Ashtekar:2011ni} that the effective dynamics incorporate the leading-order
quantum geometric effects of the full theory very well even in the deep Planck regime \cite{Singh:2009mz}.
This is especially true for the states that are sharply peaked on a classical trajectory at late
times~\cite{Kaminski:2019qjn}.

In the homogeneous and isotropic  FLRW
Universe, it is obtained by the replacement 
$$
c \rightarrow \frac{\sin\left(\mu c\right)}{\mu},
$$
where $c$ denotes the  conjugate momentum of the area operator $p$
($\propto a^2$, where $a$ is the expansion factor of the Universe),
and the lattice spacing, $\mu$, is called the polymerization
parameter. Different choices of the parameter $\mu$ give rise to
different schemes of quantization with distinct effective
dynamics. The expectation is that this would act as an ultraviolet
regulator and mitigate the initial
singularity~\cite{Corichi:2007tf,Giovannetti:2020nte}. 

In the literature, various choices of $\mu$ have been
considered~\cite{Ashtekar:2011ni,Gan:2022oiy} \footnote{{ It should be noted that the choices of $\mu$ are purely geometric and are independent of the underlying theory. As a result, they are applicable to the Brans-Dicke theory considered in this paper. For more details, see Ref. \cite{Gan:2024rga} and references therein.}}.  In particular, in the original
scheme--the $\mu_o$ scheme \cite{Ashtekar:2003hd}, the area gap was implemented as a kinematic feature on the
comoving frame, which is equivalent to introducing a fixed lattice of
constant spacing, so that  the area $\Box_{ij}$ in the ($i,j$) plane is measured by the fiducial metric ${}^{o}q_{ab}$ defined by 
Eq.~(\ref{eq3.2}). 
However, it was found that this leads to  inconsistencies~\cite{Ashtekar:2006uz}. 
After extensive investigations in the past two decades or so (see, for example, Ref.~\cite{Gan:2022oiy} and references therein), 
it was found
that the $\bar\mu$ scheme \cite{Ashtekar:2006wn}, which is sometimes also referred to as improved dynamics,  is the unique one that leads to consistent physics
not only for the homogeneous and isotropic quantum background, but also for its
linear perturbations \cite{Ashtekar:2011ni,Li:2021mop,Li:2023dwy,Agullo:2023rqq}.
The distinguishable feature of the $\bar\mu$ scheme comes from the observation that the quantization of area should refer to physical geometries instead of the fiducial metric. In particular, the area $\Box_{ij}$ in the ($i,j$) plane should be measured by  the physical metric $q_{ab}$ defined in Eq.~(\ref{eq3.1}), instead of the fiducial metric ${}^{o}q_{ab}$ defined by Eq.~(\ref{eq3.2}). Since the physical area of the faces of an elementary cell
is $p$, and each side of $\Box_{ij}$ is $\sqrt{\Delta}$  times the edge of the elementary cell, we are led to \cite{Ashtekar:2006wn}
\bq
\lb{eq4.1}
\bar\mu^2 p = \Delta \quad \Rightarrow \quad \bar\mu = \sqrt{\frac{\Delta}{p}},
\eq
 where  $ \Delta \equiv 4\pi \sqrt{3} \gamma l_{Pl}^{2}$ is the area gap in the full theory of LQG \cite{Thiemann:2002nj}.

 When generalizing the above $\bar\mu$ scheme to the homogeneous but anisotropic cases, including the Bianchi I models, in the literature
there exist two different schemes  \cite{Chiou:2006qq,Chiou:2007sp,Chiou:2007mg,Ashtekar:2009vc}, which we shall introduce below and refer them to as the $\bar\mu_A$ and $\bar\mu_B$ schemes, respectively. Both schemes have been studied in great details \cite{Martin-Benito:2009xaf,Ashtekar:2009um,Corichi:2009pp,Garay:2010sk,Wilson-Ewing:2010lkm,Gupt:2011jh,Singh:2013ava,Tarrio:2013ija,Linsefors:2014tna,
Corichi:2015ala,Wilson-Ewing:2017vju,Agullo:2020uii,Agullo:2020iqv,McNamara:2022dmf,Motaharfar:2023hil}, and showed that, among other interesting features, the singularities are resolved and replaced by quantum bounces.  More recently~ \cite{Motaharfar:2023hil}, 
it was found that the $\bar\mu_A$ quantization suffers from a  problem in which one of the triad components and associated polymerized term retains Planckian character even at large volumes. As a result, not only is the anisotropic shear not preserved across the bounce, but also the Universe can exhibit an unexpected cyclic evolution. 

\subsection{The \texorpdfstring{$\bar{\mu}_A$}{muA} quantization scheme}

The $\bar{\mu}_A$ scheme,  proposed in Ref.~\cite{Chiou:2006qq}, for the homogeneous and anisotropic Bianchi I model is
specified by
\begin{eqnarray}
\label{scheme}
\bar{\mu}_I&=&\sqrt{\frac{\Delta}{p_I}}.
\end{eqnarray}
Clearly, when   $p_I \gg \Delta$, we have
$\bar\mu_I \rightarrow 0$, and expect that the quantum effects are  very
small and the classical limit is obtained. However, near the singular
point, $p_I \simeq 0$, we have $\bar\mu_I \gg 1$, such that the quantum
effects  are expected to be very large. In the following, we shall
consider whether such effects are larger enough so that the
spacetime singularity that used to appear classically at $p_I  = 0$ will now be
regulated in the current setup. 

Returning to our model under investigation, from
Eq.~\eqref{FinalHamiltonian}, the quantum deformed Hamiltonian
constraint becomes
\begin{widetext}
\begin{eqnarray}
\label{Ceff_final}
C_{\rm eff}  &=& -\frac{1}{\gamma^2 \phi}
\left[\frac{\sin(\bar{\mu}_1\tilde{c}_1)}{\bar{\mu}_1}p_1\frac{\sin(\bar{\mu}_2\tilde{c}_2)}{\bar{\mu}_2}p_2
  +
  \frac{\sin(\bar{\mu}_2\tilde{c}_2)}{\bar{\mu}_2}p_2\frac{\sin(\bar{\mu}_3\tilde{c}_3)}{\bar{\mu}_3}p_3+
  \frac{\sin(\bar{\mu}_3\tilde{c}_3)}{\bar{\mu}_3}p_3\frac{\sin(\bar{\mu}_1\tilde{c}_1)}{\bar{\mu}_1}p_1
  \right]  \nonumber \\ 
  && +  \frac{1}{\beta \gamma^2\phi} \left[
  \frac{\sin(\bar{\mu}_1\tilde{c}_1)}{\bar{\mu}_1}p_1+
  \frac{\sin(\bar{\mu}_2\tilde{c}_2)}{\bar{\mu}_2}p_2 +
  \frac{\sin(\bar{\mu}_3\tilde{c}_3)}{\bar{\mu}_3}p_3 +  \gamma
  \pi_{\phi}\phi \right]^2   +\frac{P_{\Phi}^2}{2}.  
\end{eqnarray}
\end{widetext}
Note that, in order to study the effective theory of loop quantum Brans-Dicke
cosmology, we also want to know the effect of matter fields on the
dynamical evolution.  Hence, we have included in
Eq.~\eqref{Ceff_final} an extra massless scalar matter field $\Phi$, {with momentum conjugate $P_{\Phi}$},
and the corresponding  energy density is given by 
\bq
\lb{rho}
\rho =\frac{P_{\Phi}^2}{2p_1p_2p_3}.  
\eq
Since $C_{\rm eff}$ does not depend explicitly on the scalar field $\Phi$, we have 
\bq
\lb{rhoB}
P_{\Phi}' = - \frac{\partial C_{\rm eff}}{\partial \Phi} = 0,
\eq
that is, $P_{\Phi}$ is a constant of motion. On the other hand, the other quantum corrected
Hamilton's equations for the symmetry reduced connection parametrized
by $\tilde{c}$ with the $\bar{\mu}_A$ scheme given by
Eq.~\eqref{scheme} are given by
\begin{widetext}
\begin{eqnarray}
\label{cQuantum}
\tilde{c}_I'  &=& -\frac{1}{\gamma\phi} \left[\frac{3\sin(\bar{\mu}_I
    \tilde{c}_I)}{2\bar{\mu}_I} - \frac{c_I \cos(\bar{\mu}_I
    \tilde{c}_I)}{2} \right] \left\{ \frac{\sin(\bar{\mu}_J
  \tilde{c}_J)}{\bar{\mu}_J}p_J+\frac{\sin(\bar{\mu}_K
  \tilde{c}_K)}{\bar{\mu}_K}p_K\right.  \nonumber\\ &&-
\left.\frac{2}{\beta} \left[
  \frac{\sin(\bar{\mu}_I\tilde{c}_I)}{\bar{\mu}_I}p_I
  +\frac{\sin(\bar{\mu}_J\tilde{c}_J)}{\bar{\mu}_J}p_J
  +\frac{\sin(\bar{\mu}_K\tilde{c}_K)}{\bar{\mu}_K}p_K  +  \gamma
  \pi_{\phi}\phi \right]\right\} \nonumber \\ && + \gamma p_J p_K\left(
\rho+p_I\frac{\partial\rho}{\partial p_I}\right).
\end{eqnarray}
\end{widetext}
{}For other choices of schemes for $\bar\mu_I$, the change will be in the
terms inside the first square brackets.
Similarly, quantum corrected equations of motion for the symmetry
reduced densetized triads parametrized by $p_I$ are given by
\begin{eqnarray}
\label{pQuantum}
p_I' &=&  \frac{p_I}{\phi \gamma}  \cos(\bar{\mu}_I \tilde{c}_I)
\left\{  \frac{\sin(\bar{\mu}_J \tilde{c}_J)}{\bar{\mu}_J}p_J+
\frac{\sin(\bar{\mu}_K \tilde{c}_K)}{\bar{\mu}_K}p_K \right.
\nonumber\\ && - \left.  \frac{2}{\beta}  \left[
  \frac{\sin(\bar{\mu}_I\tilde{c}_I)}{\bar{\mu}_I}p_I+\frac{\sin(\bar{\mu}_J\tilde{c}_J)}{\bar{\mu}_J}p_J
  \right.\right.  \nonumber \\ && +
  \left. \left. \frac{\sin(\bar{\mu}_K\tilde{c}_K)}{\bar{\mu}_K}p_K +
  \gamma \pi_{\phi}\phi\right] \right\}.
\end{eqnarray}
{}Finally, the equations of motion for $\phi$ and its conjugate
momentum $\pi_\phi$ read
\begin{eqnarray}
\label{dphidt_final}
\phi'&=& \frac{2}{\gamma\beta} \Bigl[ \frac{\sin (\tilde{c}_1
    \bar{\mu}_1)}{\bar{\mu}_1}p_1 + \frac{\sin (\tilde{c}_2
    \bar{\mu}_2)}{\bar{\mu}_2}p_2 + \frac{\sin (\tilde{c}_3
    \bar{\mu}_3)}{\bar{\mu}_3}p_3 \nonumber \\ &+&  \gamma\pi_\phi
  \phi \Bigr],\\
\label{dpiphidt_final}
\pi_\phi'&=& - \frac{1}{\gamma^2\phi^2} \left\{
\frac{\sin(\bar{\mu}_1\tilde{c}_1)}{\bar{\mu}_1}p_1\frac{\sin(\bar{\mu}_2\tilde{c}_2)}{\bar{\mu}_2}p_2
\right.  \nonumber \\ &+& \left.
\frac{\sin(\bar{\mu}_2\tilde{c}_2)}{\bar{\mu}_2}p_2\frac{\sin(\bar{\mu}_3\tilde{c}_3)}{\bar{\mu}_3}p_3
+
\frac{\sin(\bar{\mu}_1\tilde{c}_1)}{\bar{\mu}_1}p_1\frac{\sin(\bar{\mu}_3\tilde{c}_3)}{\bar{\mu}_3}
p_3   \right.  \nonumber\\ &+& \left. \frac{
  (\gamma\pi_{\phi}\phi)^2}{\beta} \right.  \nonumber \\ &-&\left.
\frac{1}{\beta}\left( \frac{\sin (\tilde{c}_1
  \bar{\mu}_1)}{\bar{\mu}_1}p_1 + \frac{\sin (\tilde{c}_2
  \bar{\mu}_2)}{\bar{\mu}_2}p_2 + \frac{\sin (\tilde{c}_3
  \bar{\mu}_3)}{\bar{\mu}_3}p_3 \right)^2  \right\}.  \nonumber \\
\end{eqnarray}

The set of Eqs.~\eqref{cQuantum}-\eqref{dpiphidt_final} completely
define the quantum corrected dynamics of 
the Bianchi-I spacetime  in $\bar{\mu}_A$
scheme for BDT in the Jordan's frame.  

Once the above equations are solved,  we can compute the average Hubble parameter
$H$ and the expansion factor $a$ for the effective quantum Bianchi-I spacetime
\begin{eqnarray}
\lb{eq4.10}
H \equiv \frac{H_1+H_2+H_3}{3},\quad a \equiv \sqrt{a_1a_2a_3},
\end{eqnarray}
where    $H_I=\dot{a}_I/a_I$ and $p_I=a_Ja_Kl_0^2$ for nonrepeated indexes
$I,\, J,\, K$.  In particular, the average Hubble
parameter reads
\begin{eqnarray}
H &=& \frac{1}{6\gamma \phi\sqrt{p_1p_2p_3}} \left\{\left[
  \frac{\sin(\bar{\mu}_2 \tilde{c}_2)}{\bar{\mu}_2}p_2  +
  \frac{\sin(\bar{\mu}_3 \tilde{c}_3)}{\bar{\mu}_3}p_3 \right]\right.\nb\\
  && \times 
\cos(\bar{\mu}_1 \tilde{c}_1)  \nonumber \\ 
&&+ \left.  \left[
  \frac{\sin(\bar{\mu}_1 \tilde{c}_1)}{\bar{\mu}_1}p_1  +
  \frac{\sin(\bar{\mu}_3 \tilde{c}_3)}{\bar{\mu}_3}p_3 \right]
\cos(\bar{\mu}_2 \tilde{c}_2) \right.  \nonumber\\ && +\left.\left[
  \frac{\sin(\bar{\mu}_2 \tilde{c}_2)}{\bar{\mu}_2}p_2  +
  \frac{\sin(\bar{\mu}_1 \tilde{c}_1)}{\bar{\mu}_1}p_1  \right]
\cos(\bar{\mu}_3 \tilde{c}_3)  \right\} \nonumber \\ && -
\frac{\dot{\phi}}{6\phi \sqrt{p_1p_2p_3}}  
\left[\cos(\bar{\mu}_1 \tilde{c}_1) +\cos(\bar{\mu}_2
  \tilde{c}_2)+\cos(\bar{\mu}_3 \tilde{c}_3) \right], \nonumber \\
\label{HubbleParameter}
\end{eqnarray}
where the factor $\sqrt{p_1p_2p_3}$ comes from Eq.~\eqref{gauge}.  
To see how classical singularities can be avoided in the framework of
LQG,   
let us provide some heuristic arguments, by following the analysis provided  in ~\cite{Chiou:2007sp}. 
{}For more details, see our numerical analyses to be presented in the next section.

Defining
\begin{eqnarray}\label{GI}
\mathcal{G}_I(t')=p_I\frac{\sin(\bar{\mu}_I\tilde{c}_I)}{\bar{\mu}_I},    
\end{eqnarray}
and using Eqs.~\eqref{cQuantum}-\eqref{pQuantum}, one can show that
\begin{eqnarray}\label{dGIdtp}
\mathcal{G}_I'(t') = \gamma \cos(\bar{\mu}_I\tilde{c}_I) p_1p_2p_3
\left(\rho+p_I\frac{\partial \rho}{\partial p_I}\right).
\end{eqnarray}
Now setting the quantum effective constraint, Eq.~\eqref{Ceff_final},
to zero, one obtains
\begin{eqnarray}
\lb{eq4.14}
 \gamma^2  p_1p_2p_3 \rho_{\rm eff} = \mathcal{G}_1\mathcal{G}_2  +
 \mathcal{G}_2\mathcal{G}_3 + \mathcal{G}_1\mathcal{G}_3,  
\end{eqnarray}
where we have used Eq.~\eqref{dphidt_final} and defined the effective
energy density as
\begin{eqnarray}\label{rhoeff}
\rho_{\rm eff}=\frac{\beta\dot{\phi}^2}{4} + \rho\phi.    
\end{eqnarray}
Since $\rho$ is given by Eq.~(\ref{rho}), we find the identity
\bq
\lb{4.16}
\rho+p_I\frac{\partial
  \rho}{\partial p_I} = 0,
  \eq
which means that $\mathcal{G}_I$ are constants.  {}For constant
$\mathcal{G}_I$, from Eq.~\eqref{GI}, given that the sines have a
maximum value of unit, it sets a lower bound on $p_I$,
\begin{eqnarray}\label{pI_bound}
p_I\geq(|\mathcal{G}|\sqrt{\Delta})^{2/3}.    
\end{eqnarray}
This is to say that the results of Ref.~\cite{Chiou:2007sp} also hold
in the present case of BDT, provided that $\rho\to\rho_{\rm eff}$.
Then Eq.~\eqref{pI_bound} shows that all scale factors are bounded
from below, which proves that a nonsingular behavior is present.

\subsection{The \texorpdfstring{$\bar{\mu}_B$}{muB} quantization scheme}

In the $\bar{\mu}_B$   scheme, we have
~\cite{Tarrio:2013ija}
\begin{eqnarray}\label{alternative}
\bar{\mu}_I = \sqrt{\frac{\Delta\, p_I}{p_J p_K}},
\end{eqnarray}
where $I$, $J,$ and $K$ are nonrepeated indexes. The corresponding effective
Hamiltonian takes the same form as that given by Eq.~(\ref{Ceff_final}). {}Following the same steps provided in the last subsection,
we can obtain the corresponding Hamiltonian equations. In particular, it can be shown that
\begin{eqnarray}\label{phi2}
\dot{\phi}^2 &=& \frac{4}{\gamma^2\beta^2(p_1p_2p_3)} \left[ \frac{p_1
    \sin (\tilde{c}_1 \bar{\mu}_1)}{\bar{\mu}_1} + \frac{p_2 \sin
    (\tilde{c}_2 \bar{\mu}_2)}{\bar{\mu}_2} \right.  \nonumber \\ &+&
  \left.  \frac{p_3 \sin (\tilde{c}_3 \bar{\mu}_3)}{\bar{\mu}_3} +
  \gamma\pi_\phi \phi \right]^2,
\end{eqnarray}
which is now written with respect to the cosmic time $t$.  Now setting
once again the quantum effective Hamiltonian constraint,
Eq.~\eqref{Ceff_final}, to zero, one obtains
\begin{widetext}
\begin{eqnarray}
\label{Heff0}
&-&\frac{1}{ \gamma^2 \phi}
  \left[\frac{\sin(\bar{\mu}_1\tilde{c}_1)}{\bar{\mu}_1}p_1\frac{\sin(\bar{\mu}_2\tilde{c}_2)}{\bar{\mu}_2}p_2
    +
    \frac{\sin(\bar{\mu}_2\tilde{c}_2)}{\bar{\mu}_2}p_2\frac{\sin(\bar{\mu}_3\tilde{c}_3)}{\bar{\mu}_3}p_3+
    \frac{\sin(\bar{\mu}_3\tilde{c}_3)}{\bar{\mu}_3}p_3\frac{\sin(\bar{\mu}_1\tilde{c}_1)}{\bar{\mu}_1}p_1
    \right]  \nonumber \\ &+&   \frac{1}{\beta \gamma^2\phi} \left[
    \frac{\sin(\bar{\mu}_1\tilde{c}_1)}{\bar{\mu}_1}p_1+
    \frac{\sin(\bar{\mu}_2\tilde{c}_2)}{\bar{\mu}_2}p_2 +
    \frac{\sin(\bar{\mu}_3\tilde{c}_3)}{\bar{\mu}_3}p_3 + \gamma
    \pi_{\phi}\phi \right]^2   + p_1p_2p_3\rho=0.  
\end{eqnarray}
\end{widetext}
Note that in the second line of the above equation the term in the
square brackets can be written in terms of Eq.~\eqref{phi2}, then one
obtains
\begin{eqnarray}\label{sin_rhoeff}
\frac{\beta\dot{\phi}^2}{4 \phi}   + \rho &=&\frac{1}{\gamma^2 \phi}
\left[
  \frac{\sin(\bar{\mu}_1\tilde{c}_1)\sin(\bar{\mu}_2\tilde{c}_2)}{\bar{\mu}_1\bar{\mu}_2p_3}
  \right.  \nonumber \\ && + \left.
  \frac{\sin(\bar{\mu}_2\tilde{c}_2)\sin(\bar{\mu}_3\tilde{c}_3)}{\bar{\mu}_2\bar{\mu}_3p_1}
  +
  \frac{\sin(\bar{\mu}_3\tilde{c}_3)\sin(\bar{\mu}_1\tilde{c}_1)}{\bar{\mu}_1\bar{\mu}_3p_2}
  \right] .   \nonumber \\
\end{eqnarray}
At this point, we apply the aforementioned alternative scheme, given
by Eq.~\eqref{alternative}, in  Eq.~\eqref{sin_rhoeff}, and  define an effective energy density
$\rho_{\rm eff}$ as
\begin{eqnarray}\label{rhoeffrhoc}
\frac{\rho_{\rm eff}}{\rho_c} &\equiv&   \frac{\gamma^2 \Delta}{3}
\left(\frac{\beta\dot{\phi}^2}{4}   + \rho\phi\right) \nonumber \\ &=&
\frac{1}{3}\Bigl[\sin(\bar{\mu}_1\tilde{c}_1)\sin(\bar{\mu}_2\tilde{c}_2)
  + \sin(\bar{\mu}_2\tilde{c}_2)\sin(\bar{\mu}_3\tilde{c}_3)
  \nonumber \\ &&+
  \sin(\bar{\mu}_1\tilde{c}_1)\sin(\bar{\mu}_3\tilde{c}_3)\Bigr].  
\end{eqnarray}
Clearly, such defined effective energy density  is bounded from above by its maximal
value $\rho_c$, given by $\rho_{c} = 3/(\gamma^2 \Delta)$.
These results are in agreement with Ref.~\cite{Zhang:2016twe} in the
isotropic limit.

Now, a maximal value of energy density implies 
{a minimum value of spatial volume}.  Therefore, initializing the Universe with a
contracting phase, a quantum bounce is predicted in this scenario.  This
confirms the removal of an initial singularity in the framework of LQG
owing to the fact that geometry is quantized and there exists a
nonzero minimal area gap.   Starting with an initially contracting phase,
in a classical setup, the collapse will inevitably lead to a spacetime singularity.  Thus, once again,
through this analysis we conclude that quantum geometric effects at
the Planck scale provide repulsive forces  and replace classical spacetime 
singularities by  nonsingular quantum bounces.

{}For completeness, let us now write a closed form of the modified Friedmann
equation and that can be derived using the current scheme.  From
Eqs.~\eqref{HubbleParameter} and~\eqref{phi2}, we obtain that
\begin{widetext}
\begin{eqnarray}\label{Friedmann_attempt}
\left(H+\frac{\dot{\phi}}{2\phi}\right)^2    &=&
\left\{\frac{1}{6\gamma\sqrt{\Delta}\phi} \left[ \sin (\tilde{c}_1
  \bar{\mu}_1)  \cos (\tilde{c}_2 \bar{\mu}_2) + \sin (\tilde{c}_2
  \bar{\mu}_2) \cos (\tilde{c}_1 \bar{\mu}_1) + \sin (\tilde{c}_1
  \bar{\mu}_1) \cos (\tilde{c}_3 \bar{\mu}_3) + \sin (\tilde{c}_3
  \bar{\mu}_3) \cos (\tilde{c}_1 \bar{\mu}_1) 
  \right.\right.\nonumber\\ &+& \left.\left.  \sin (\tilde{c}_2
  \bar{\mu}_2) \cos (\tilde{c}_3 \bar{\mu}_3) + \sin (\tilde{c}_3
  \bar{\mu}_3) \cos (\tilde{c}_2 \bar{\mu}_2) \right] +
\frac{\dot{\phi}}{2 \phi} \left[ 1 - \frac{ \cos (\tilde{c}_1
    \bar{\mu}_1) + \cos (\tilde{c}_2 \bar{\mu}_2) + \cos (\tilde{c}_3
    \bar{\mu}_3)} {3}  \right]\right\}^2.
\end{eqnarray}
Inspired by Eq.~\eqref{rhoeffrhoc}, we define the directional
effective energy density in the I direction as
\begin{eqnarray}
\rho_{\rm eff}^{(I)} = \frac{3\sin^2(\tilde{c}_I \bar{\mu}_I)}{
  \gamma^2 \Delta}. 
\end{eqnarray}
Using this definition, Eq.~\eqref{Friedmann_attempt} can now be
expressed as
\begin{eqnarray}\label{Friedmann_attemptv2}
\left(H+\frac{\dot{\phi}}{2\phi}\right)^2    &=& \left\{
\frac{1}{6\phi} \left[ \sqrt{ \frac{1}{3}\rho_{\rm eff}^{(1)}
    \left(1-\frac{\rho_{\rm eff}^{(2)}}{\rho_c}\right) } +
  \sqrt{\frac{1}{3}\rho_{\rm eff}^{(2)} \left(1-\frac{\rho_{\rm
        eff}^{(1)}}{\rho_c}\right) } + \sqrt{\frac{1}{3}\rho_{\rm
      eff}^{(1)} \left(1-\frac{\rho_{\rm eff}^{(3)}}{\rho_c}\right)} 
  \right.\right.\nonumber\\ &+& \left.  \sqrt{\frac{1}{3}\rho_{\rm
      eff}^{(3)} \left(1-\frac{\rho_{\rm eff}^{(1)}}{\rho_c}\right)} +
  \sqrt{\frac{1}{3}\rho_{\rm eff}^{(2)} \left(1-\frac{\rho_{\rm
        eff}^{(3)}}{\rho_c}\right)} + \sqrt{\frac{1}{3}\rho_{\rm
      eff}^{(3)} \left(1-\frac{\rho_{\rm eff}^{(2)}}{\rho_c}\right)}
  \right] \nonumber\\ &+& \left.  \frac{\dot{\phi}}{2 \phi} \left[
  1-\frac{1}{3}\left( \sqrt{1-\frac{\rho_{\rm eff}^{(1)}}{\rho_c}} +
  \sqrt{1-\frac{\rho_{\rm eff}^{(2)}}{\rho_c}} +
  \sqrt{1-\frac{\rho_{\rm eff}^{(3)}}{\rho_c}} \right)
  \right]\right\}^2. 
\end{eqnarray}
\end{widetext}
The above equation reduces to the isotropic BD
theory~\cite{Zhang:2016twe} in the limit where the directional
effective energy densities are equal.  Also, the effective {}Friedmann
equation of LQC is recovered for $\phi=1$, when $\rho_{\rm
  eff}\to\rho$. 

\section{Dynamics of Bianchi I Universe} 
\label{dynamics}

In this section we present our numerical results for the effective dynamics of
the models considered in the last section. We show the evolution of the
scale factors and Hubble parameters of the Bianchi I Universe for  LQC  and  BDT  in each of the quantization schemes.
To be able to compare our results with the ones obtained in Ref.~\cite{Motaharfar:2023hil} in the framework of LQC,
here we choose the same initial conditions as those used by the authors of that reference. On the other hand, 
in the framework of BDT, we choose the 
{initial condition for the triad and the connection variables as close as possible to those in LQC}. 
In particular,  in all of our numerical results for LQC, we choose the initial conditions $c_1=-0.13$, $c_2=-0.12$, $p_1=10^3$, $p_2=2\times10^3$, $p_3=3 \times10^3$, while $c_3$ is determined from the Hamiltonian constraint, Eq.~\eqref{Heff0}, with $(\phi,
\dot\phi, \rho) = (1, 0, 0)$. 

With the above in mind, in {}Figs.~\ref{a_I_A} and~\ref{a_I_B} we show our numerical results for the directional scale factors $a_I \;(I=1,2,3)$ of the Bianchi-I  spacetimes, in both LQC and BDT, for the
$\bar{\mu}_A$ and $\bar{\mu}_B$ schemes, given,  respectively by Eqs.~\eqref{scheme} and~\eqref{alternative},  with the parameters described in the figure captions.   
 Each figure has four panels for each scheme. Figures 1(a)  and 2(a) shows the results for the LQC model, while Figs. 1(b)-1(d) and 2(b)-2(d) show the results for the BDT models.
Notice that we use initial conditions for $\dot{\phi}$ instead of $\pi_\phi$ because it is more transparent when considering the limit  $\mathrm{BDT}\to\mathrm{LQC}$, which happens for $(\phi,\dot{\phi})\to(1,0)$.

\begin{widetext}

\begin{center}
\begin{figure}[htb!]
\subfigure[]{\includegraphics[width=0.47\textwidth]{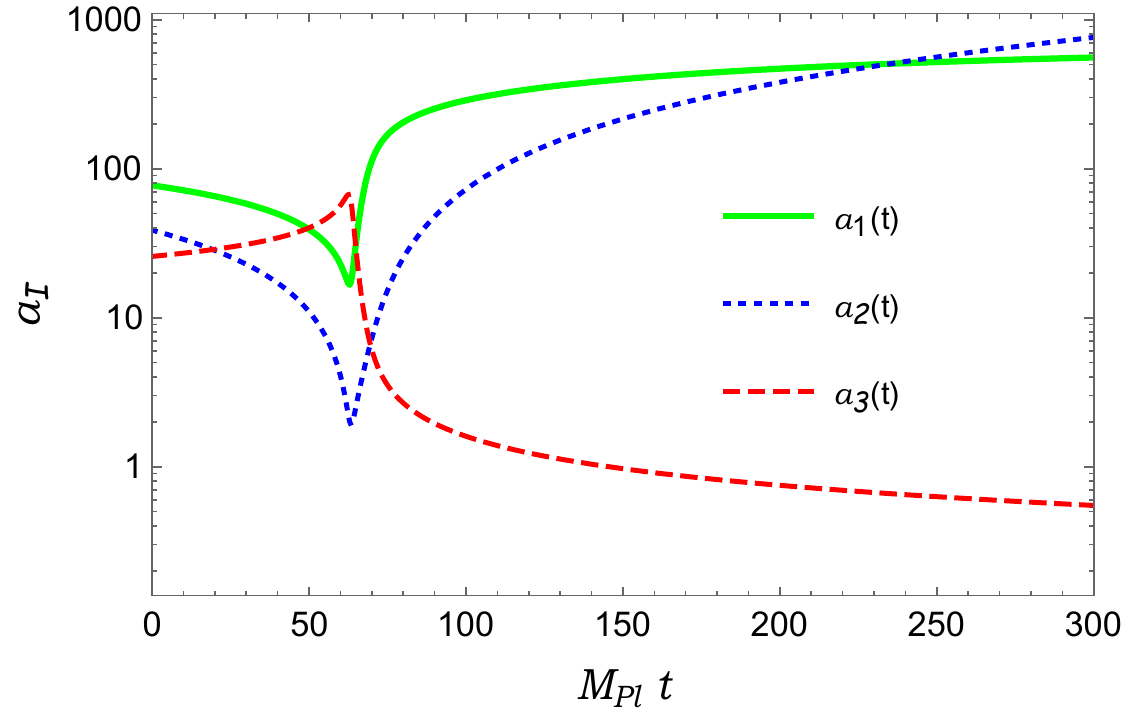}}
\subfigure[]{\includegraphics[width=0.47\textwidth]{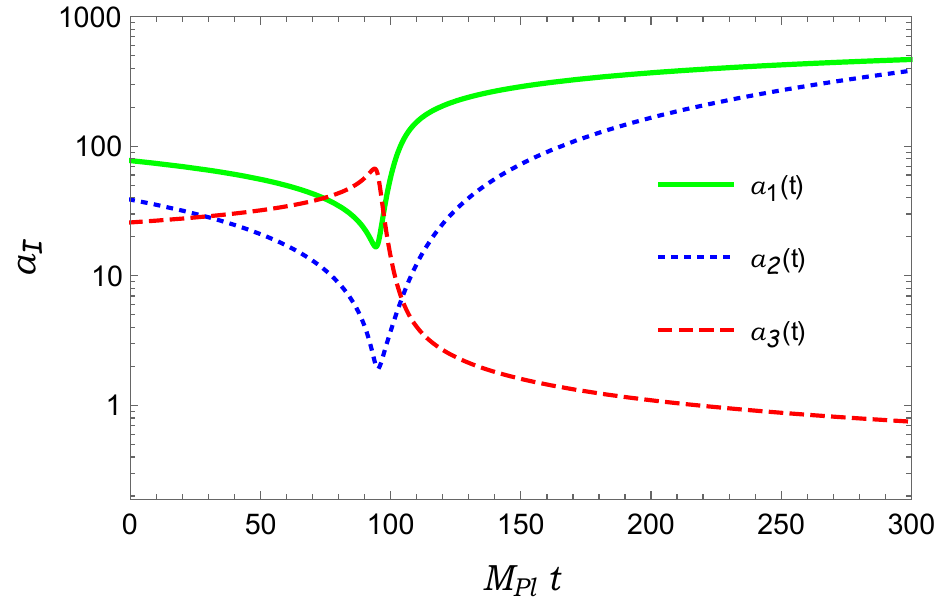}}\\
\subfigure[]{\includegraphics[width=0.47\textwidth]{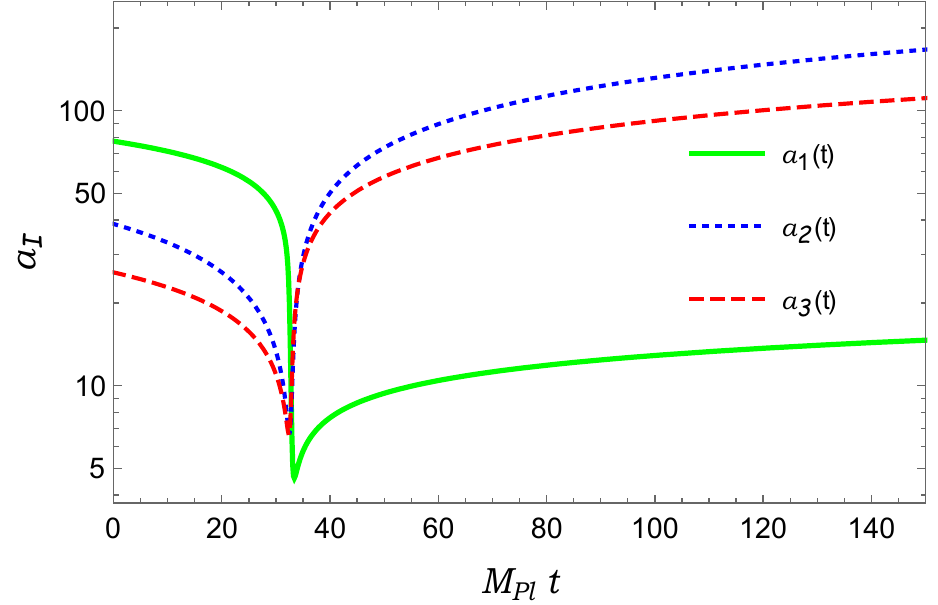}}
\subfigure[]{\includegraphics[width=0.47\textwidth]{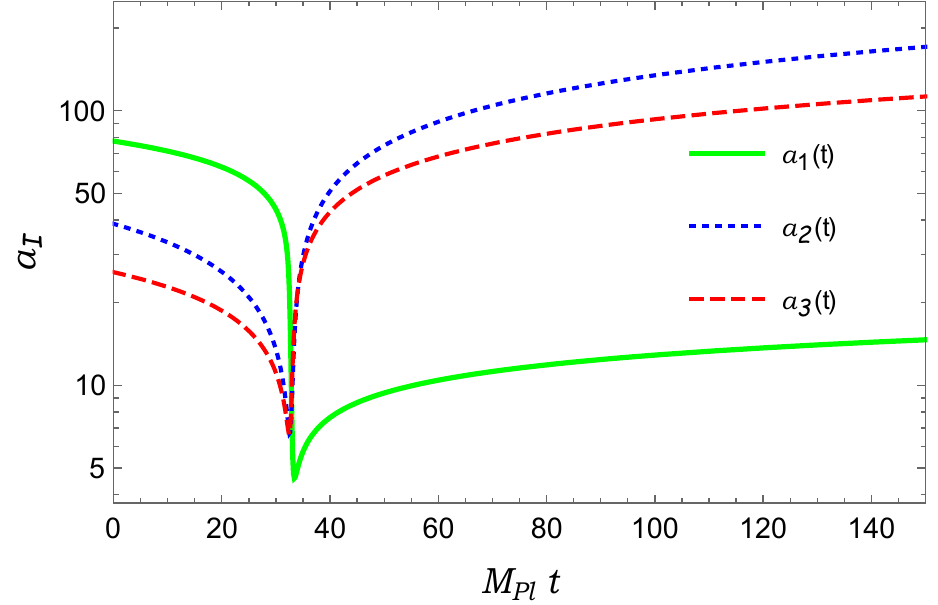}}
\caption{The evolution of the directional scale factors $a_I \; (I=1,2,3)$ of the Bianchi-I spacetimes, where  (a) is for LQC  and  (b) - (d) are for BDT. The initial conditions for LQC are $c_1=-0.13$, $c_2=-0.12$, $p_1=10^3$, $p_2=2\times10^3$, $p_3=3\times10^3$. 
{}For BDT, we adopt the same initial conditions as those of LQC with the addition of
$(\phi,\dot{\phi})=(1.5,0)$ for (b), $(\phi,\dot{\phi})=(1.0,5.5\times10^{-5})$ for Panel (c), and
$(\phi,\dot{\phi})=(1.0,-5.5\times10^{-5})$ for  (d), respectively, and set $\omega=2\times10^5$ \cite{Zhang:2017sym} for the BD parameter in all these three last. All plots correspond to the $\bar{\mu}_A$ scheme.}
\label{a_I_A}
\end{figure}
\end{center}

\begin{center}
\begin{figure}[htb!]
\subfigure[]{\includegraphics[width=0.47\textwidth]{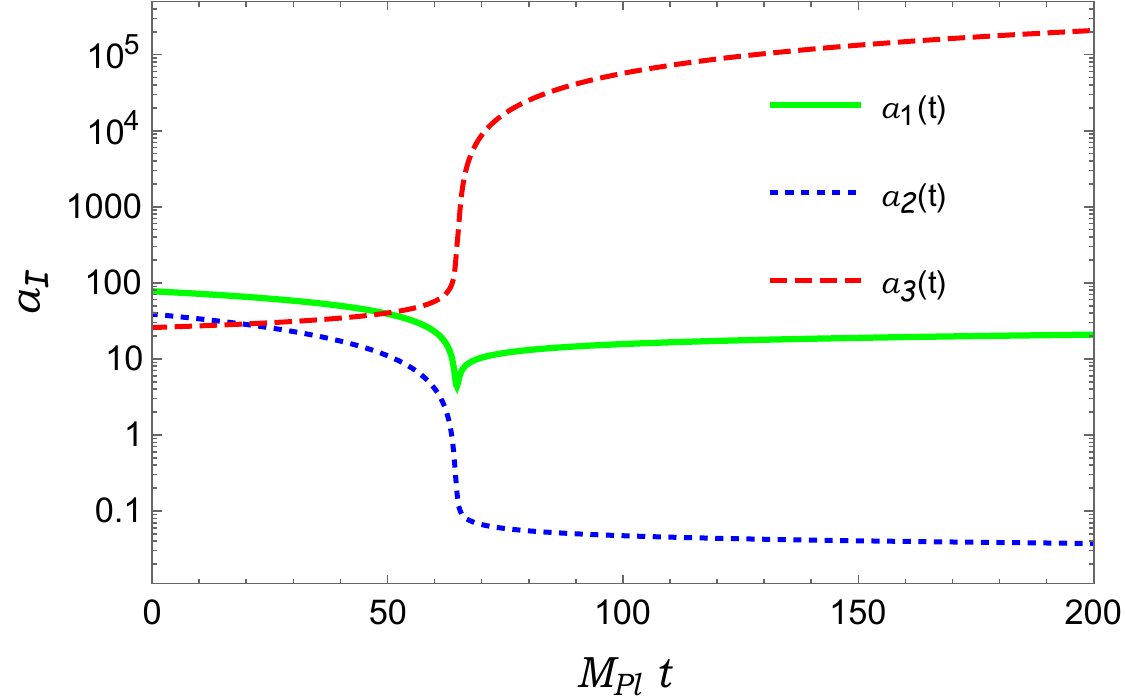}}
\subfigure[]{\includegraphics[width=0.47\textwidth]{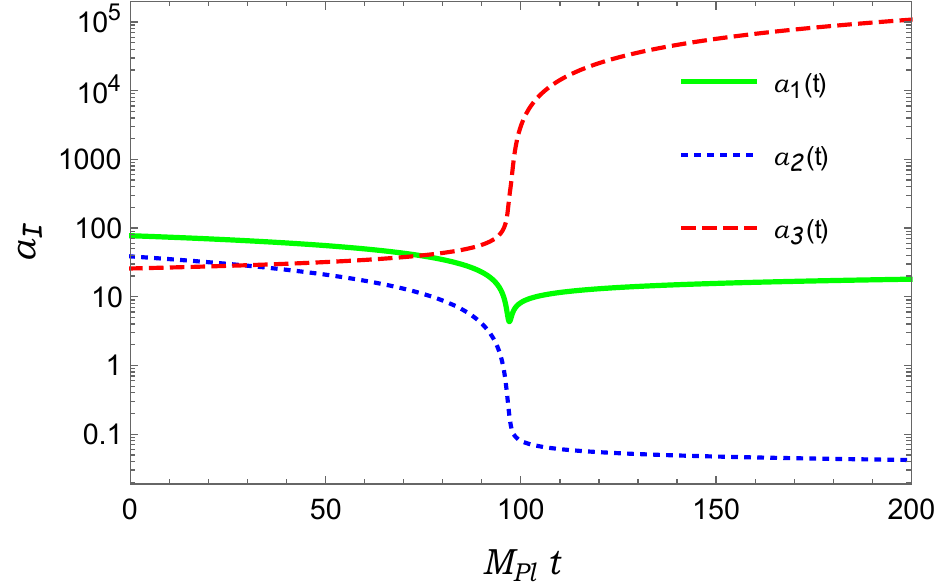}}\\
\subfigure[]{\includegraphics[width=0.47\textwidth]{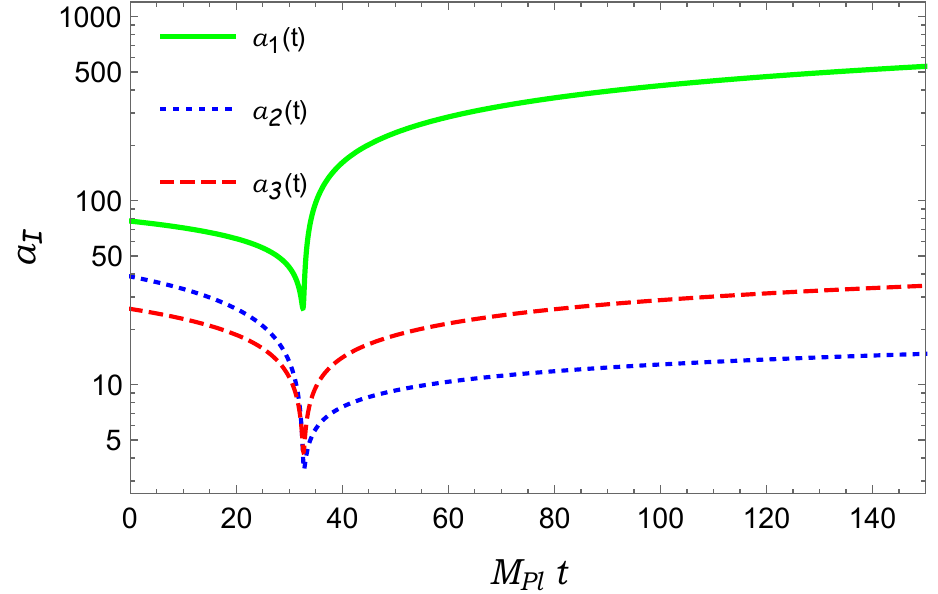}}
\subfigure[]{\includegraphics[width=0.47\textwidth]{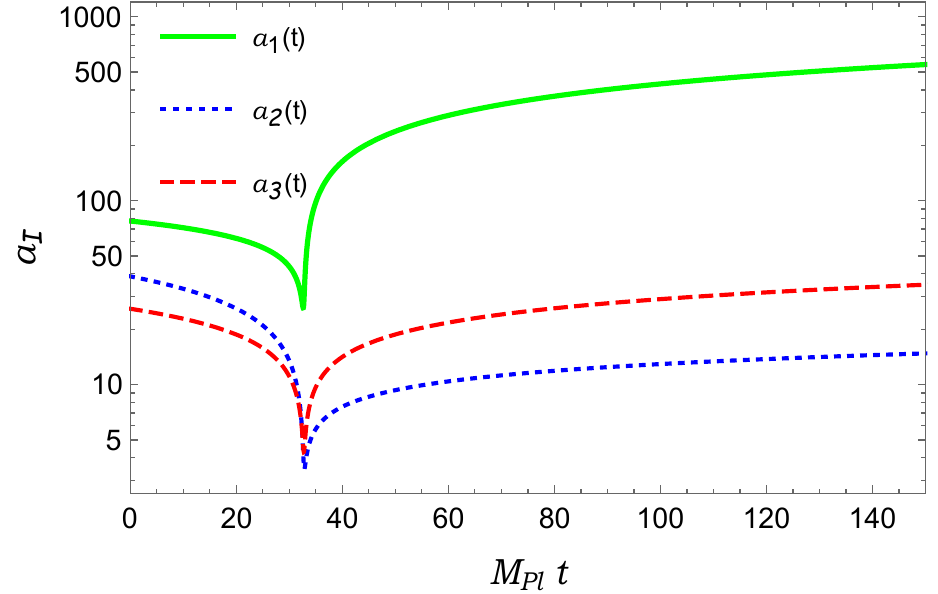}}
\caption{The same as in {}Fig.~\ref{a_I_A}, but for the $\bar{\mu}_B$ scheme.}
\label{a_I_B}
\end{figure}
\end{center}

\end{widetext}

In {}Figs.~\ref{a_I_A}(a) and~\ref{a_I_B}(a) we show the results for the directional scale
factors $a_I \; (I=1,2,3)$ of the Bianchi-I spacetime in the framework of  LQC  for the
$\bar{\mu}_A$ and $\bar{\mu}_B$ schemes, respectively. 
Comparing them with the corresponding ones given in Ref.~\cite{Motaharfar:2023hil}, one can see clearly the consistency among these figures within the numerical errors. As a result,   the numerical code used in the current paper and the one used in 
Ref.~\cite{Motaharfar:2023hil} are  all reliable and give the same results.

Then, comparing {}Fig.~\ref{a_I_A}(a) with {}Fig.~\ref{a_I_A}(b), we can see that the deviations of BDT from that of LQC are simply the delay 
of the bounce in each direction, while the main qualitative behavior is preserved. 
This is because in the present case the initial condition of $\dot\phi$ is chosen to be zero, as that in LQC, so at the beginning the BDT effects are negligible. This type of dynamics
is consistent with a Kasner-like phase, i.e., expansion followed by
contraction in one direction and contraction followed by expansion in the others.  
It is important to note that even though there
appears contraction in one of the three directions, we have checked explicitly that it never reaches zero (e.g., reaches a spacetime singularity) in that direction, but it always reaches asymptotically a nonzero value. Similar behavior has been also observed previously 
in LQC, see, for example, Refs.~\cite{Chiou:2007sp}~\cite{Chiou:2007mg} ~\cite{Motaharfar:2023hil}.
From the numerical
studies that we have performed, we have also verified that the inclusion of a massless scalar field $\Phi$
does not change the results significantly, as long as $\dot\phi(t_i) = 0$, where $t_i$ denotes the initial time.

However, from {}Fig.~\ref{a_I_A}(c) we can see that once
the initial condition of $\dot\phi$ is different from zero, the deviations become significant. In particular, all the three directional scale factors are increasing after the bounce (at $M_{\rm Pl} t_B \simeq 33$) and soon after the bounce they become comparable with their initial values. This is sharply in contrast to the LQC case (first seen in Ref.~\cite{Motaharfar:2023hil}), in which one of the three directional scale factors is always decreasing, and only two of them are increasing after the bounce. Here we see that this is also true for 
both $\mu_A$ and $\mu_B$ schemes when $\dot \phi=0$, as one can see from {}Figs.~\ref{a_I_A}(b) and \ref{a_I_B}(b). On the other hand, the difference between {}Figs.~\ref{a_I_A}(c) and \ref{a_I_A}(d) is the choice of the sign of $\dot\phi$ at the initial moment. In fact, in 
{}Fig.~\ref{a_I_A}(c) we chose $(\phi,\dot{\phi})=(1.0, 5.5\times10^{-5})$, while in {}Fig.~\ref{a_I_A}(d) we chose $(\phi,\dot{\phi})=(1.0, -5.5\times10^{-5})$. The results for both cases show that the directional scale
factors remain the same. This can be seen from Eq.~(\ref{eq4.14}), in which its right-hand side  is a constant, while its left-hand side 
is independent of the signs of $\dot\phi$, as one can see from Eq.~(\ref{rhoeff}). Recall that $p_I \equiv l_0^2\, a_Ja_K$ for nonrepeated indexes
$I,\, J,\, K$.
Similar results as the ones for the $\mu_A$ scheme are also seen for the $\mu_B$ schemes.  
Comparing {}Fig.~\ref{a_I_B}(a) with  {}Fig.~\ref{a_I_B}(b), we can see that the directional scale factors behave quite similar, modulated  the simple delay of the bounce in each direction. In particular,
only two of the three directional scale factors are increasing after the bounce, while the third one is always decreasing, which is again consistent as the results observed in Ref.~\cite{Motaharfar:2023hil}.  Again, this is because of the choice of the initial condition $\dot\phi(t_i) = 0$, which is initially identical to those in the LQC case. However, once $\dot\phi(t_i) \neq 0$,
as shown explicitly in {}Figs.~\ref{a_I_B}(c) and \ref{a_I_B}(d), the deviations become large. Remarkably, all the three directional scale factors
are increasing right after the bounce and soon become comparable with their initial values. That is, if the evolution of the spacetime starts from a classical point, it will collapse to a minimal but nonzero volume, where the quantum bounce occurs and replaces the classical spacetime singularity by a finite quantum bounce. Afterwards, the spacetime starts to expand in each of the three directions, and soon approaches to a size that is comparable with the  initial one. This is true not only in terms of volumes and areas but also in terms of the size in each of the three directions. This is 
dramatically different from that observed in LQC for the $\mu_A$ scheme~\cite{Motaharfar:2023hil}. 

\begin{figure}[!htb]
\centering  
\subfigure[]{\includegraphics[width=0.47\textwidth]{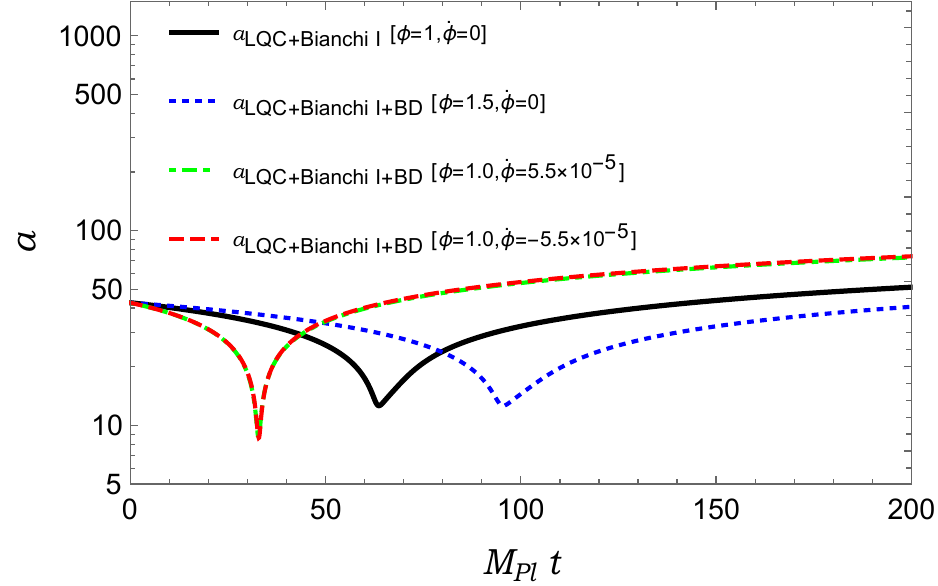}}
\subfigure[]{\includegraphics[width=0.47\textwidth]{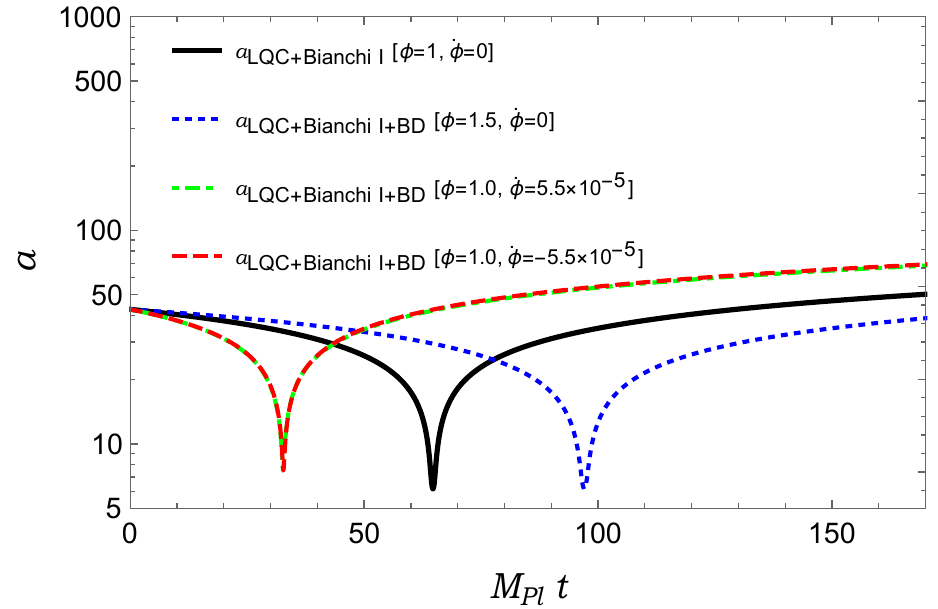}}
\caption{The {average} scale factor $a$ of the Bianchi I spacetime for the $\bar{\mu}_A$ [(a)] and $\bar{\mu}_B$ [(b)] schemes, where  the LQC model is presented by the black curve and the BDT models are presented by  the blue, yellow and red curves, respectively for the choices,  $(\phi,\dot{\phi})=(1.5,0)$,
$(\phi,\dot{\phi})=(1.0,5.5\times10^{-5})$, and  $(\phi,\dot{\phi})=(1.0,-5.5\times10^{-5})$. The other initial conditions are the same as those given in {}Fig.~\ref{a_I_A}
for the $\mu_A$ scheme, and those given in Fig. \ref{a_I_B}
for the $\mu_B$ scheme.}
\label{a}
\end{figure}

\begin{figure}[!htb]
\centering  
\subfigure[]{\includegraphics[width=0.47\textwidth]{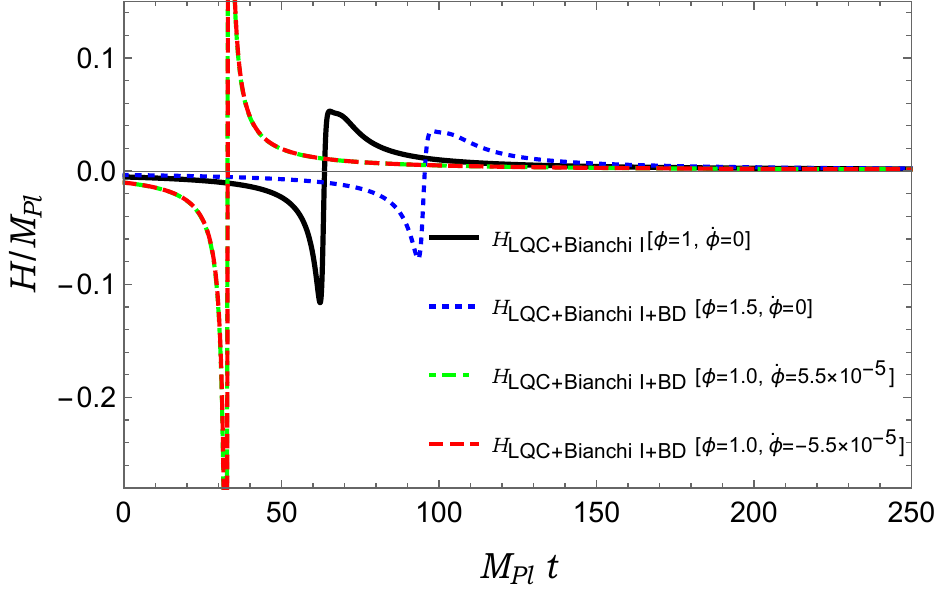}}
\subfigure[]{\includegraphics[width=0.47\textwidth]{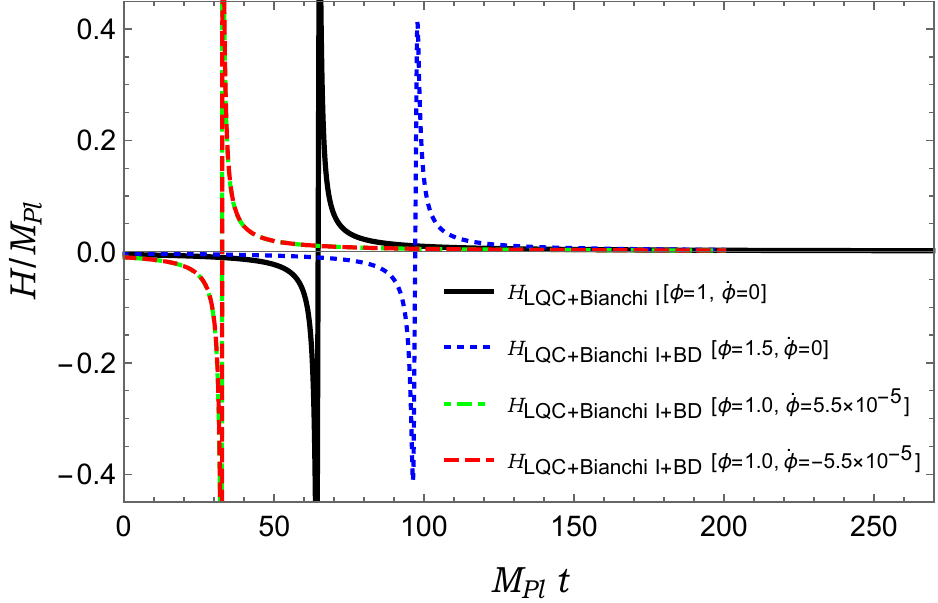}}
\caption{
The {average} Hubble parameters $H$ of the Bianchi I spacetime for the $\bar{\mu}_A$ scheme [(a)] and the $\bar{\mu}_B$ scheme [(b)], where  the LQC model is presented by the black curve and the BDT models are presented by  the blue, yellow and red curves, respectively, for the choices,  $(\phi,\dot{\phi})=(1.5,0)$,
$(\phi,\dot{\phi})=(1.0,5.5\times10^{-5})$, and  $(\phi,\dot{\phi})=(1.0,-5.5\times10^{-5})$. The other initial conditions are the same as those given in Fig. \ref{a_I_A}
for the $\mu_A$ scheme and Fig. \ref{a_I_B} for the $\mu_B$ scheme. 
}
\label{H}
\end{figure}

To understand the properties of the models further, in {}Figs.~\ref{a} and ~\ref{H}, we show the numerical results for the average scale factor $a$ and Hubble
parameter $H$, defined by Eq.~(\ref{eq4.10}), for both of the $\mu_A$ and $\mu_B$ schemes. In each of these figures, the initial conditions are chosen the same as the 
corresponding ones given in {}Figs. \ref{a_I_A} and \ref{a_I_B}, respectively, for the $\mu_A$ and $\mu_B$ schemes. In particular, {}Figs. ~\ref{a} and \ref{H} show similar behaviors for the LQC models as presented in  Refs.~\cite{Chiou:2007mg,Motaharfar:2023hil}, while for the BDT models, the effects of the BD scalar $\phi$ depend on the choice of the initial conditions. When $\dot\phi(t_i) = 0$, the effects with respect to LQC are simply to delay the bounce, and make the increase of the scale factor slower, while for
the initial conditions with $\dot\phi(t_i) \neq 0$, the effects are opposite, that is, the bounce happens earlier, and the average scale factor increases faster after the bounce. Again, such behaviors do not depend on the signs of $\dot\phi(t_i)$ for the same reasons as explained above for the directional scale factors $a_I$'s. Hence, the results for the same absolute value of $\dot\phi(t_i)$ are overlapping.

Note that an overall behavior of the average Hubble parameter is also consistent with the
results obtained earlier in Ref.~\cite{Artymowski:2013qua}, although the effects of the initial conditions on the shapes of $a$ and $H$ were not considered there.

In addition, in all the figures we have plotted the curves only up to $M_{\rm Pl} t \lesssim 300$, but we have found that the asymptotical behavior of the curves  remains the same. In fact, we have run our codes up to $M_{\rm Pl} t \simeq 10^6$ and found no changes. So, their asymptotical behavior remains the same.

Analysis of these figures leads us to conclude that, for the representative parameters and initial conditions considered, all results consistently demonstrate that the classical singularity is replaced by a regular quantum bounce. In the absence of BD effects, this bounce occurs around $t= 65/M_{\rm Pl}$. However, the BD effects can shift the bounce to earlier or later times depending on the initial conditions of the BD scalar $\phi(t_i)$ and its derivative $\phi(t_i)$, where $t_i$ denotes the initial time. 
Furthermore, the BD effects prevent the suppression of the directional scale factors after the bounce, a stark contrast to what was previously seen in the LQC case~\cite{Chiou:2007mg,Motaharfar:2023hil}.

{}Finally, we note that we have explicitly verified that the values of the
parameters and  initial conditions  chosen are quite generic in the
sense that, by changing them to other reasonable values, we obtain similar behaviors, so our main  conclusions and results
 are robust with different initial conditions. Also, in this section we do not include plots for the cases in which the massless scalar field $\Phi$ (which is equivalent to a  stiff fluid) is present,
as we find that  it does not change the general results presented
here, and the qualitative behavior  also remains the
same. 

\section{Conclusions}\label{DR}

In this paper we have derived the quantum-corrected effective
background dynamics for BDT in the Bianchi I model in the Jordan's
frame in the framework of LQG.  We have cast the theory in terms of
the Yang-Mills phase space variables suitable for loop quantization
and we have presented the quantum-corrected effective dynamics that
resolve the spacetime singularity.  The quantization is made by
considering two different schemes, as given by Eqs.~\eqref{scheme} and
\eqref{alternative}, respectively.  The former set provides lower
bounds on the scale factor, Eq.~\eqref{pI_bound}, proving the
existence of a nonsingular quantum behavior of spacetime.  On the
other hand, the latter scheme gives us an upper bound to the effective
energy density, i.e., the critical density $\rho_c$, implying a
minimal value of spatial volume.  
Additionally, we have also derived
the quantum  corrected Friedmann equation for this model inspired by
the idea of a directional energy density.  We have explicitly shown
that the effective dynamics of the Bianchi I Universe must undergo a
smooth transition from a contracting phase to a expanding phase
through the quantum bounce.  Our analytical results, therefore, show
robustness of the singularity resolution and the consistency of the
effective dynamics beyond general relativity. 

In addition, we have also found different results obtained in  BDT from those obtained in LQC. In particular, we have found that  
all the three directional scale factors are generically increasing after the quantum bounce and soon reach to their initial values, as can be seen 
from  {}Figs.~\ref{a_I_A}(c) and  \ref{a_I_A}(d) for the $\mu_A$ scheme and in {}Figs.~\ref{a_I_B}(c) and  \ref{a_I_B}(d) for the $\mu_B$ scheme. This is sharply in contrast to the LQC case, in which one of the three directional scale factors is always decreasing after the bounce, and only two of them are increasing, as can be seen from {}Figs.~\ref{a_I_A}(a) and  \ref{a_I_B}(a) in both schemes.

As a perspective of future study in the context of the model considered here, it would be of interest to explore any possible observational
features it might lead and that can be detected  in the forthcoming
cosmic microwave background experiments. An extension of the analysis, like the one performed
in earlier studies in the context of LQC for homogeneous and isotropic
models~\cite{Agullo:2013ai,Zhu:2017jew,Benetti:2019kgw,Barboza:2022hng,Barboza:2020jux},
would be of great interest. 

Likewise, the analysis of the perturbations in
the model presented here and the possibility of an inflationary phase
following the quantum bounce could be eventually done.  Concerning any
possible observational signatures in  LQC scenarios with an
anisotropic background, one can mention, for instance, the work of
Ref.~\cite{Agullo:2022klq}, where an analysis of the possible
observational features in such models was performed.  Hence, an
extension of such analysis of perturbations in the context of the BDT
anisotropic LQC will be necessary in order to elaborate the observable
predictions of the model. 

Issues related to particle production due to
the bounce can also be
important~\cite{Graef:2020qwe,Quintin:2014oea,Tavakoli:2014mra,Haro:2015zda,Celani:2016cwm,Scardua:2018omf,Hipolito-Ricaldi:2016kqq,Graef:2017nyv,SVicente:2022ebm}
and their analysis and consequences in the context of anisotropic loop
quantum BDT would be worthwhile to be pursued in the future as well. 

\begin{acknowledgments}
This work was partly done when M.S. was visiting the Inter-University
Centre for Astronomy and Astrophysics (IUCAA), Pune, India. M.S. would
like to express his sincere gratitude to Professor Dr. Naresh Dadhich
for his kind invitation to visit IUCAA in the fall of 2022. The
vibrant research environment and the scenic beauty of the campus was
an impetus to carry this piece of work.  L.L.G is supported by research grants from Conselho Nacional de
Desenvolvimento Cient\'{\i}fico e Tecnol\'ogico (CNPq), Grant
No. 307636/2023-2 and from the
Fundacao Carlos Chagas Filho de Amparo a Pesquisa do Estado do Rio de
Janeiro (FAPERJ), Grant No.  E-26/201.297/2021.  L.L.G. would like to
thank IPMU--Kavli Institute for the Physics and Mathematics of the
Universe for warm hospitality and also thank Professor Elisa Ferreira for
the constant scientific inspiring discussions.  R.O.R. acknowledges
financial support  by research grants from Conselho Nacional de
Desenvolvimento Cient\'{\i}fico e Tecnol\'ogico (CNPq), Grant
No. 307286/2021-5, and from Funda\c{c}\~ao Carlos Chagas Filho de
Amparo \`a Pesquisa do Estado do Rio de Janeiro (FAPERJ), Grant
No. E-26/201.150/2021.   A.W. is partially supported by the US NSF
Grant No. PHY2308845.
\end{acknowledgments}



\begin{thebibliography}{999}

\bibitem{Ashtekar:2021kfp}
A. Ashtekar and E. Bianchi, A short review of loop quantum gravity, Rep. Prog. Phys. 84, 042001 (2021).
\bibitem{Thiemann:2007pyv}
T. Thiemann, Modern Canonical Quantum General Relativity (Cambridge University Press, Cambridge, England, 2007), ISBN 978-0-511-75568-2, 978-0-521-84263-1,
10.1017/CBO9780511755682.
\bibitem{Rovelli:2008zza}
 C. Rovelli, Loop quantum gravity, Living Rev. Relativity 11, 5 (2008).
\bibitem{Bojowald:2010qpa}
 M. Bojowald, Canonical Gravity and ApplicationsCosmology, Black Holes, and Quantum Gravity (Cambridge University Press, Cambridge, England, 2010), ISBN 978-0-521-19575-1, 978-0-511-98513-3.
\bibitem{Gambini:2011zz}
R. Gambini and J. Pullin, A First Course in Loop Quantum
Gravity (Oxford University Press, Oxford, 2011).
\bibitem{Rovelli:2014ssa}
C. Rovelli and F. Vidotto, Covariant Loop Quantum
Gravity: An Elementary Introduction to Quantum Gravity and Spinfoam Theory (Cambridge University Press,
Cambridge, England, 2014), ISBN 978-1-107-06962-6,
978-1-316-14729-0.
\bibitem{Thiemann:2002nj}
 T. Thiemann, Lectures on loop quantum gravity, Lect. Notes Phys. 631, 41 (2003).
\bibitem{Ashtekar:2004eh}
A. Ashtekar and J. Lewandowski, Background independent quantum gravity: A status report, Classical Quantum Gravity 21, R53 (2004).
\bibitem{Rovelli:2004tv}
C. Rovelli, Quantum Gravity (University Press, New York, 2004), 10.1017/CBO9780511755804.
\bibitem{Ashtekar:1994wa}
 A. Ashtekar and J. Lewandowski, Differential geometry on the space of connections via graphs and projective limits, J. Geom. Phys. 17, 191 (1995)
 
\bibitem{Ashtekar:1997fb}
A. Ashtekar and J. Lewandowski, Quantum theory of
geometry II. Volume operators, Adv. Theor. Math. Phys. 1, 388 (1998).
\bibitem{Ashtekar:2011ni}
 A. Ashtekar and P. Singh, Loop quantum cosmology: A
status report, Classical Quantum Gravity 28, 213001 (2011).
\bibitem{Li:2021mop}
 B. F. Li, P. Singh, and A. Wang, Phenomenological implications of modified loop cosmologies: An overview, Front. Astron. Space Sci. 8, 701417 (2021).
\bibitem{Barca:2021qdn}
G. Barca, E. Giovannetti, and G. Montani, An overview on the nature of the bounce in LQC and PQM, Universe 7, 327 (2021). 
\bibitem{Li:2023dwy}
B. F. Li and P. Singh, Loop quantum cosmology: Physics of singularity resolution and its implications,\\ \textit{arXiv:2304.05426}.
\bibitem{Agullo:2023rqq}
 I. Agullo, A. Wang, and E. Wilson-Ewing, Loop quantum cosmology: Relation between theory and observations, arXiv:2301.10215.
\bibitem{Bianchi:2008es}
E. Bianchi, The length operator in loop quantum gravity, Nucl. Phys. B807, 591 (2009).
\bibitem{Ma:2010fy}
Y. Ma, C. Soo, and J. Yang, New length operator for loop quantum gravity, Phys. Rev. D 81, 124026 (2010).

\bibitem{Bojowald:2005epg}
M. Bojowald, Loop quantum cosmology, Living Rev.
Relativity 8, 11 (2005).
\bibitem{Brandenberger:2012zb}
R. H. Brandenberger, The matter bounce alternative to
inflationary cosmology, arXiv:1206.4196.
\bibitem{Lehners:2008vx}
J. L. Lehners, Ekpyrotic and cyclic cosmology, Phys. Rep. 465, 223 (2008).
\bibitem{Cai:2013vm}
Y. F. Cai, R. Brandenberger, and P. Peter, Anisotropy in a nonsingular bounce, Classical Quantum Gravity 30, 075019 (2013).
\bibitem{Sharma:2018vnv}
 M. Sharma, M. Shahalam, Q. Wu, and A. Wang, Preinflationary dynamics in loop quantum cosmology: Monodromy potential, J. Cosmol. Astropart. Phys. 11 (2018) 003.
\bibitem{Jin:2018wdx}
W. J. Jin, Y. Ma, and T. Zhu, Pre-inflationary dynamics of Starobinsky inflation and its generalization in loop quantum Brans-Dicke cosmology, J. Cosmol. Astropart. Phys. 02 (2019) 010.
\bibitem{Li:2018fco}
B. F. Li, P. Singh, and A. Wang, Qualitative dynamics and inflationary attractors in loop cosmology, Phys. Rev. D 98, 066016 (2018).
\bibitem{Levy:2024naz}
G. L. L. W. Levy and R. O. Ramos, $\alpha$-attractor potentials in loop quantum cosmology, Phys. Rev. D 110, 043507 (2024).
\bibitem{Shahalam:2017wba}
M. Shahalam, M. Sharma, Q. Wu, and A. Wang, Preinflationary dynamics in loop quantum cosmology: Power-law potentials, Phys. Rev. D 96, 123533 (2017).

\bibitem{Chiou:2006qq}
D. W. Chiou, Loop quantum cosmology in Bianchi type I
models: Analytical investigation, Phys. Rev. D 75, 024029 (2007).
\bibitem{Chiou:2007sp}
 D. W. Chiou and K. Vandersloot, The Behavior of nonlinear anisotropies in bouncing Bianchi I models of loop quantum cosmology, Phys. Rev. D 76, 084015 (2007).
\bibitem{Chiou:2007mg}
D. W. Chiou, Effective dynamics, big bounces and scaling symmetry in Bianchi type I loop quantum cosmology, Phys. Rev. D 76, 124037 (2007).
\bibitem{Martin-Benito:2008dfr}
M. Martin-Benito, G. A. Mena Marugan, and \\ T. Pawlowski, 
Loop quantization of vacuum Bianchi I cosmology, 
Phys. Rev. D 78, 064008 (2008).


\bibitem{Ashtekar:2009vc}
A. Ashtekar and E. Wilson-Ewing, Loop quantum cosmology of Bianchi I models, Phys. Rev. D 79, 083535 (2009).
\bibitem{Martin-Benito:2009xaf}
 M. Martin-Benito, G. A. M. Marugan, and T. Pawlowski, Physical evolution in loop quantum cosmology: The example of vacuum Bianchi I, Phys. Rev. D 80, 084038 (2009).
\bibitem{Ashtekar:2009um}
A. Ashtekar and E. Wilson-Ewing, Loop quantum cosmology of Bianchi type II models, Phys. Rev. D 80, 123532 (2009).
\bibitem{Corichi:2009pp}
 A. Corichi and P. Singh, A geometric perspective on
singularity resolution and uniqueness in loop quantum
cosmology, Phys. Rev. D 80, 044024 (2009).
\bibitem{Garay:2010sk}
L. J. Garay, M. Martin-Benito, and G. A. Mena Marugan, Inhomogeneous loop quantum cosmology: Hybrid quantization of the Gowdy model, Phys. Rev. D 82, 044048 (2010).
\bibitem{Wilson-Ewing:2010lkm}
E. Wilson-Ewing, Loop quantum cosmology of Bianchi
type IX models, Phys. Rev. D 82, 043508 (2010).
\bibitem{Gupt:2011jh}
B. Gupt and P. Singh, Contrasting features of anisotropic loop quantum cosmologies: The role of spatial curvature, Phys. Rev. D 85, 044011 (2012).
\bibitem{Singh:2013ava}
P. Singh and E. Wilson-Ewing, Quantization ambiguities and bounds on geometric scalars in anisotropic loop quantum cosmology, Classical Quantum Gravity 31, 035010 (2014).
\bibitem{Tarrio:2013ija}
P. Tarrío, M. F. M´endez, and G. A. M. Marugán, Singularity avoidance in the hybrid quantization of the Gowdy model, Phys. Rev. D 88, 084050 (2013).
\bibitem{Linsefors:2014tna}
L. Linsefors and A. Barrau, Exhaustive investigation of the duration of inflation in effective anisotropic loop quantum cosmology, Classical Quantum Gravity 32, 035010 (2015).
\bibitem{Corichi:2015ala}
A. Corichi and E. Montoya, Loop quantum cosmology of
Bianchi IX: Effective dynamics, Classical Quantum Gravity 34, 054001 (2017).
\bibitem{Wilson-Ewing:2017vju}
E. Wilson-Ewing, The loop quantum cosmology bounce as
a Kasner transition, Classical Quantum Gravity 35, 065005 (2018).
\bibitem{Agullo:2020uii}
 I. Agullo, J. Olmedo, and V. Sreenath, Hamiltonian theory
of classical and quantum gauge invariant perturbations in
Bianchi I spacetimes, Phys. Rev. D 101, 123531 (2020).
\bibitem{Agullo:2020iqv}
I. Agullo, J. Olmedo, and V. Sreenath, Observational
consequences of Bianchi I spacetimes in loop quantum
cosmology, Phys. Rev. D 102, 043523 (2020).
\bibitem{McNamara:2022dmf}
A. M. McNamara, S. Saini, and P. Singh, Novel relationship between shear and energy density at the bounce in nonsingular Bianchi I spacetimes, Phys. Rev. D 107, 026003 (2023).
\bibitem{Motaharfar:2023hil}
M. Motaharfar, P. Singh, and E. Thareja, Classicality and uniqueness in the loop quantization of Bianchi I spacetimes, Phys. Rev. D 109, 086013 (2024).
\bibitem{Belinsky:1970ew}
V. A. Belinsky, I. M. Khalatnikov, and E. M. Lifshitz, Oscillatory approach to a singular point in the relativistic cosmology, Adv. Phys. 19, 525 (1970).
\bibitem{Aluri:2022hzs}
P. K. Aluri, P. Cea, P. Chingangbam, M. C. Chu, R. G.
Clowes, D. Hutsem´ekers, J. P. Kochappan, A. M. Lopez, L. Liu, N. C. M. Martens et al., Is the observable Universe consistent with the cosmological principle?, Classical Quantum Gravity 40, 094001 (2023).
\bibitem{Yeung:2022smn}
 S. Yeung and M. C. Chu, Directional variations of cosmological parameters from the Planck CMB data, Phys. Rev. D 105, 083508 (2022).
\bibitem{Agullo:2022klq}
I. Agullo, J. Olmedo, and E. Wilson-Ewing, Observational constraints on anisotropies for bouncing alternatives to inflation, J. Cosmol. Astropart. Phys. 10 (2022) 045.
\bibitem{Amirhashchi:2017mur}
H. Amirhashchi, Probing dark energy in the scope of a
Bianchi type I spacetime, Phys. Rev. D 97, 063515 (2018).
\bibitem{Zhang:2011vg}
X. Zhang and Y. Ma, Nonperturbative loop quantization of scalar-tensor theories of gravity, Phys. Rev. D 84, 104045 (2011).
\bibitem{Artymowski:2013qua}
 M. Artymowski, Y. Ma, and X. Zhang, Comparison between Jordan and Einstein frames of Brans-Dicke gravity a la loop quantum cosmology, Phys. Rev. D 88, 104010 (2013).
\bibitem{Han:2015jsa}
 Y. Han, K. Giesel, and Y. Ma, Manifestly gauge invariant perturbations of scalar-tensor theories of gravity, Classical Quantum Gravity 32, 135006 (2015).
\bibitem{Han:2019mvj}
Y. Han, Loop quantum cosmological dynamics of scalar tensor theory in the Jordan frame, Phys. Rev. D 100, 123541 (2019).
\bibitem{Song:2020pqm}
S. Song, C. Zhang, and Y. Ma, Alternative dynamics in loop quantum Brans-Dicke cosmology, Phys. Rev. D 102, 024024 (2020).
\bibitem{Zhang:2017sym}
X. Zhang, J. Yu, T. Liu, W. Zhao, and A. Wang, Testing Brans-Dicke gravity using the Einstein telescope, Phys. Rev. D 95, 124008 (2017).
\bibitem{Kiefer:2004xyv}
C. Kiefer, Quantum Gravity, 3rd ed. (Oxford University Press, England, 2012).
\bibitem{DeWitt:2007mi}
B. S. DeWitt and G. Esposito, An introduction to quantum gravity, Int. J. Geom. Methods Mod. Phys. 05, 101 (2008).
\bibitem{Baez:1995sj}
J. C. Baez and J. P. Muniain, Gauge Fields, Knots and
Gravity (World Scientific Publishing Company, Singapore, 1994), Vol. 4.

\bibitem{Olmo:2011fh} 


G. J. Olmo and H. Sanchis-Alepuz, Hamiltonian formulation of Palatini f(R) theories a la Brans-Dicke, Phys. Rev. D 83, 104036 (2011).








\bibitem{Jantzen:2001me}
R. T. Jantzen, Spatially homogeneous dynamics: A unified picture, arXiv:gr-qc/0102035.
\bibitem{Taveras:2008ke}
V. Taveras, Corrections to the Friedmann equations from LQG for a Universe with a free scalar field, Phys. Rev. D 78, 064072 (2008).
\bibitem{Singh:2009mz}
P. Singh, Are loop quantum cosmos never singular?,
Classical Quantum Gravity 26, 125005 (2009).
\bibitem{Kaminski:2019qjn}
W. Kaminski, M. Kolanowski, and J. Lewandowski,
Dressed metric predictions revisited, Classical Quantum Gravity 37, 095001 (2020).
\bibitem{Corichi:2007tf}
A. Corichi, T. Vukasinac, and J. A. Zapata, Polymer
quantum mechanics and its continuum limit, Phys. Rev.
D 76, 044016 (2007).
\bibitem{Giovannetti:2020nte}
E. Giovannetti, G. Barca, F. Mandini, and G. Montani,
Polymer dynamics of isotropic universe in ashtekar and in volume variables, Universe 8, 302 (2022).
\bibitem{Gan:2022oiy}
W. C. Gan, X. M. Kuang, Z. H. Yang, Y. Gong, A. Wang,
and B. Wang, Nonexistence of quantum black and white
hole horizons in an improved dynamic approach, Sci. China Phys. Mech. Astron. 67, 280411 (2024).

\bibitem{Gan:2024rga}
W.-C. Gan and A. Wang, A new quantization scheme of
black holes in effective loop quantum gravity, Phys. Rev. D  111, 026017 (2025).
\bibitem{Ashtekar:2003hd}
A. Ashtekar, M. Bojowald, and J. Lewandowski, Mathematical structure of loop quantum cosmology, Adv. Theor. Math. Phys. 7, 233 (2003).
\bibitem{Ashtekar:2006uz}
A. Ashtekar, T. Pawlowski, and P. Singh, Quantum nature of the big bang: An analytical and numerical investigation, Phys. Rev. D 73, 124038 (2006).
\bibitem{Ashtekar:2006wn}
A. Ashtekar, T. Pawlowski, and P. Singh, Quantum nature of the big bang: Improved dynamics, Phys. Rev. D 74, 084003 (2006).
\bibitem{Zhang:2016twe}
X. Zhang, Loop quantum cosmology, modified gravity and extra dimensions, Universe 2, 15 (2016).
\bibitem{Agullo:2013ai}
I. Agullo, A. Ashtekar, and W. Nelson, The preinflationary dynamics of loop quantum cosmology: Confronting quantum gravity with observations, Classical Quantum Gravity 30, 085014 (2013).
\bibitem{Zhu:2017jew}
T. Zhu, A. Wang, G. Cleaver, K. Kirsten, and Q. Sheng, Pre-inflationary universe in loop quantum cosmology, Phys. Rev. D 96, 083520 (2017).
\bibitem{Benetti:2019kgw}
 M. Benetti, L. Graef, and R. O. Ramos, Observational
constraints on warm inflation in loop quantum cosmology, J. Cosmol. Astropart. Phys. 10 (2019) 066.
\bibitem{Barboza:2022hng}
L. N. Barboza, G. L. L. W. Levy, L. L. Graef, and R. O. Ramos, Constraining the Barbero-Immirzi parameter from the duration of inflation in loop quantum cosmology, Phys. Rev. D 106, 103535 (2022).
\bibitem{Barboza:2020jux}
L. N. Barboza, L. L. Graef, and R. O. Ramos, Warm
bounce in loop quantum cosmology and the prediction
for the duration of inflation, Phys. Rev. D 102, 103521 (2020).
\bibitem{Graef:2020qwe}
L. L. Graef, R. O. Ramos, and G. S. Vicente, Gravitational particle production in loop quantum cosmology, Phys. Rev. D 102, 043518 (2020).
\bibitem{Quintin:2014oea}
 J. Quintin, Y. F. Cai, and R. H. Brandenberger, Matter creation in a nonsingular bouncing cosmology, Phys. Rev. D 90, 063507 (2014).
\bibitem{Tavakoli:2014mra}
Y. Tavakoli and J. C. Fabris, Creation of particles in a cyclic universe driven by loop quantum cosmology, Int. J. Mod. Phys. D 24, 1550062 (2015).
\bibitem{Haro:2015zda}
 J. Haro and E. Elizalde, Gravitational particle production in bouncing cosmologies, J. Cosmol. Astropart. Phys. 10 (2015) 028.
\bibitem{Celani:2016cwm}
 D. C. F. Celani, N. Pinto-Neto, and S. D. P. Vitenti, Particle creation in bouncing cosmologies, Phys. Rev. D 95, 023523 (2017).

\bibitem{Scardua:2018omf}
 A. Scardua, L. F. Guimarães, N. Pinto-Neto, and G. S. Vicente, Fermion production in bouncing cosmologies, Phys. Rev. D 98, 083505 (2018).
\bibitem{Hipolito-Ricaldi:2016kqq}
W. S. Hipolito-Ricaldi, R. Brandenberger, E. G. M. Ferreira, and L. L. Graef, Particle production in ekpyrotic scenarios, J. Cosmol. Astropart. Phys. 11 (2016) 024. 


\bibitem{Graef:2017nyv}
L.~L.~Graef, W.~S.~Hipolito-Ricaldi, E.~G.~M.~Ferreira and R.~Brandenberger,
Dynamics of cosmological perturbations and reheating in the anamorphic universe,
J. Cosmol. Astropart. Phys. 04 (2017) 004.

\bibitem{SVicente:2022ebm}
G.~S.~Vicente, R.~O.~Ramos and L.~L.~Graef,
Gravitational particle production and the validity of effective descriptions in loop quantum cosmology,
Phys. Rev. D {106}, 043518 (2022).

\end{thebibliography}
\end{document}